# Spontaneously broken Standard Model (SM) symmetries and the Goldstone theorem protect the Higgs mass and ensure that it has no Higgs Fine Tuning Problem (HFTP)

Bryan W. Lynn

Mitchell Institute for Fundamental Physics & Astronomy
Texas A&M University, College Station, Texas 77843;
University College London, London WC1E 6BT, UK; bryan.lynn@cern.ch

## Abstract

More than 40 years ago, B.W.Lee and K.Symanzik proved (but did not say it) that Ward-Takahashi identities, along with tadpole renormalization, a Vacuum Stability Condition (VSC), force all S-Matrix ultra-violet quadratic divergences (UV-QD) to be absorbed into the physical renormalized pseudo-scalar pion mass-squared $m_\pi^2$, in *O(4)* linear sigma models ( $O(4)L\Sigma M$ ) across the $\langle H \rangle$ vs. $m_\pi^2 / \lambda^2$ half-plane: e.g. the bare "Higgs" VEV $\langle H_{Bare} \rangle$ is neither UV-QD divergent nor "fine-tuned". We show that all UV-QD remnants vanish identically in the "Goldstone mode" $m_\pi^2 \to 0$ limit, which restores axial-vector current conservation $\partial_\mu \vec{A}_\mu \to 0$, a Symmetry Restoration Condition (SRC). The Goldstone-mode "Higgs" mass is protected by the *O(4)* symmetry (as realized by Higgs VSC, Goldstone SRC, spontaneous symmetry breaking (SSB) and the Goldstone theorem) and is not fine-tuned.

We insist that self-consistent renormalization of the SM requires that the scalar-sector UV-QD-corrected effective Lagrangians of the SM and Goldstone-mode $O(4)L\Sigma M$ are smoothly identical in the zero-gauge-coupling limit. Lee/Symanzik's two conditions must therefore be imposed on the SM vacuum and excited states. 1) A Higgs VSC disallows it from simply disappearing into the vacuum. 2) A Goldstone SRC governing SM Nambu-Goldstone Bosons (NGB) insists that the pre-Higgs-mechanism longitudinal $W_\mu^\pm, Z_\mu$ masses-squared $m_{\pi;NGB;SM}^2$ be exactly zero. At 1-loop, independent of regularization scheme, we show that $\langle H_{Bare} \rangle$ is neither UV-QD divergent nor fine-tuned, and that $m_{\pi;NGB;SM}^2$ absorbs all SM S-Matrix UV-QD: these vanish identically in the $m_{\pi;NGB;SM}^2 \to 0$ limit, i.e. in the SSB SM. No fine-tuning (even with near-Planck-scale UV cut-off) is necessary for a weak-scale Higgs mass. Our "Higgs no-fine-tuning theorem" is simply another (albeit un-familiar) consequence of the Goldstone theorem, an exact property of the SM vacuum and spectrum.

We show that our 1-loop SM results can (almost certainly) be extended to include all-orders perturbative electro-weak and QCD loops, so that, to all perturbative SM loop orders, no UV-QD fine-tuning is necessary for a weak-scale physical SM Higgs mass ~ 126 *GeV*.

SM symmetries, SSB and the Goldstone theorem are sufficient to protect the bare and renormalized Higgs masses, and ensure that the SM does not suffer a HFTP: it is un-necessary to impose any new Beyond the Standard-Model (BSM) symmetries. Mistaken belief in a HFTP in the 1-loop SM has historically driven an expectation that new BSM physics must appear < 14 TeV. But our results re-open the possibility that LHC discovery potential might be confined to SM physics. The crucial SM make/break test (a win/win scenario!) is LHC discovery/exclusion of the SM Higgs with its mass below the upper bound predicted by high-precision 1-electroweak-loop LEP1/SLC physics.

# 1: Introduction

In this paper, we are only interested in the classification/disposition of ultra-violet quadratic divergences (**UV-QD**), not logarithmic divergences or finite parts, arising from quantum loops in the stand-alone Standard Model (**SM**), treated as a flat-space quantum field theory:
- The SM is not embedded or integrated into any higher scale Beyond the Standard Model (**BSM**) physics;
- SM loop integrals are cut off at a short-distance UV scale $\Lambda$;
- Although the cut-off can be taken to be near the Planck scale $\Lambda \sim M_{Planck}$, quantum gravitational loops are not to be included;

This paper concerns stability and symmetry restoration protection against only UV-QD. It does not address any of the other, more usual, stability issues [1]: e.g. Landau poles, very heavy Higgs mass, strongly interacting Higgs sectors, triviality bounds, appearance of tachyons, very light Higgs mass, negative quartic-coupling instability bounds, etc. Nor does it address any explanation for the numerical values of SM parameters: e.g. the hierarchy of quark and lepton masses, etc. For pedagogical clarity, the reader might imagine we are studying the pure SM situation where *"... if the Higgs mass is between 130 and 200 GeV, this analysis does not require new physics below the Planck scale!"*, Pierre Ramond, [1, pg. 177]. For SM experimental self-consistency, the Higgs mass is taken to be below the upper limit set by 1-loop-corrected SM high precision electroweak (**HPEW**) total cross sections, forward-backward asymmetries to leptons and inclusive polarization asymmetries and high precision LEP1/SLC data.

For pedagogical simplicity, we ignore logarithmic divergences and finite contributions unnecessary to our explanation. We will not distinguish between bare fields and dimensionless couplings and their renormalized values but, for dimension-2 coefficients of relevant operators, will distinguish between renormalized values: e.g. the Higgs vacuum expectation value (**VEV**)-squared $\langle H \rangle^2$, and the UV-QD bare counter-term coefficient $\delta\mu^2$. We drop all vacuum energy/bubble contributions as beyond the scope of this paper. All SM 1-loop Feynman diagrams relevant to this paper were calculated and agreed [2,3,4,5,6,7,8,9] long ago. We refer the interested reader to that vast literature for specifics. We use Euclidean metric and the Feynman-diagram naming convention of Ref. [2].

Since the symmetries, particle content and parameter space of the SM – i.e. local and global symmetries, Higgs', Nambu-Goldstone Bosons (**NGB**), quarks, leptons, gauge bosons, ghosts, gauge couplings, Yukawas, GIM, etc. - are quite baroque, this paper examines the SM in two stages of complexity. We begin with the weak-scale un-gauged bosonic *O(4)* linear sigma model (*O(4)L$\Sigma$M*), later coupled to SM quarks and leptons. We then extend those results to the full physical 1-electroweak-loop SM case. Our 1-loop results are then extended to include all-orders perturbative electro-weak loops and all-orders perturbative QCD corrections. We focus on the specific cases of weak-scale un-gauged Goldstone-mode-*O(4)L$\Sigma$M* and the SM and show that neither theory suffers a Higgs Fine-Tuning Problem (**HFTP**).

A completely separate issue (from the HFTP) is the Weak/Gravitational Scale Hierarchy Problem (**W/GSHP**). Both weak-scale Goldstone-mode-*O(4)L$\Sigma$M* and the SM are sometimes said to suffer a W/GSHP because they are unable to predict or explain the enormous splitting between the weak scale and the next larger scale (e.g. classical gravitational physics Planck scale $M_{Planck}$). We make no attempt to address that aesthetic problem here.

Section 2 clarifies the zero-gauge-coupling limit of the SM, together with the correct renormalization of spontaneously broken un-gauged $O(4)L\Sigma M$ coupled to SM quarks and leptons. Most of the calculations (if not the effective Lagrangian presentation) in Section 2 are not new and have been common knowledge for more than four decades, but we will need these results to understand the SM case. Section 2A studies bosonic $O(4)L\Sigma M$ in the Higgs VEV $\langle H \rangle$ vs. $m_\pi^2 / \lambda^2$ half-plane. Section 2B re-calculates 1-loop UV-QD in 2-point self-energies and 1-point tadpoles. It shows that Ward-Takahashi identities, together with tadpole renormalization (i.e. a Higgs Vacuum Stability Condition (**VSC**)), force all these to be absorbed into the physical renormalized pseudo-scalar pion (pole) mass-squared $m_\pi^2$. We remind the reader that $\langle H \rangle^{Bare}$ is not UV-QD, receiving only logarithmic divergent corrections (proved to all loop orders 4 decades ago). Section 2C defines the HFTP addressed in this paper, which emerges in 1-loop-UV-QD-corrected $m_\pi^2 \neq 0$ $O(4)L\Sigma M$ and, most clearly, in the $m_\pi^2 \neq 0, \langle H \rangle = 0$ "Wigner mode" limit. Section 2D studies the opposite "Goldstone mode" limit $m_\pi^2 = 0, \langle H \rangle \neq 0$ and shows that the Symmetry Restoration Condition (SRC) embedded there causes all 1-loop UV-QD to vanish identically as the pions become Nambu-Goldstone Bosons (**NGB**) $m_\pi^2 \to m_{\pi;NGB;L\Sigma M}^2 = 0$ ( i.e. NGB masses are exactly zero!), and the theory to have no HFTP there. Section 2D also extends bosonic Goldstone-mode-$O(4)L\Sigma M$ results to all-orders in loop-perturbation theory. Section 2E further extends those Goldstone-mode bosonic $O(4)L\Sigma M$ results to include UV-QD from virtual SM quarks and leptons. Section 3A is a reminder of some necessary 1-loop SM results: especially that "oblique" loop high precision electroweak (HPEW) physics and $\langle H \rangle^{Bare}$ are not UV-QD, and receive at worst only logarithmically divergent corrections. Section 3B re-calculates all 1-loop UV-QD in the SM (i.e. including non-zero gauge couplings, gauge bosons, ghosts, Higgs mechanism, etc.) and shows that, after gauge-dependent Higgs VSC tadpole renormalization, all SM UV-QD are absorbed by NGB masses-squared $m_{\pi;NGB;SM}^2$. Section 3C shows that Lee/Symanzik's Goldstone SRC cause all UV-QD and HFTP to vanish identically in the full 1-loop spontaneously broken SM, where $m_{\pi;NGB;SM}^2 = 0$. Section 3D shows that our 1-loop SM results can (almost certainly) be extended to include all-loop-orders perturbative electroweak and QCD corrections. Section 4 re-traces our primary result to VSC tadpole renormalization, spontaneous symmetry breaking (**SSB**) and SRC enforcement of the Goldstone theorem, re-opens the possibility that the discovery potential of the LHC might be confined to SM physics, gives some phenomenological consequences and reminds the reader of the 1-loop SM make/break test at LHC. Appendix 1 shows that the vanishing of 1-loop UV-QD remnants in $O(4)L\Sigma M$ and the SM does not depend on the choice of $n$-dimensional or Pauli-Villars UV cut-off regularization. Appendix 2 discusses the "fine-tuning discontinuity" arising in Goldstone-mode un-gauged $O(4)L\Sigma M$ (or the SM) when calculations <u>begin</u> using the Goldstone-mode (or the SM) bare Lagrangian. Appendix 3 re-calculates all 1-loop SM UV-QD in a general $R_\xi$ gauge. Appendix 4 outlines B.W.Lee's 1970 proof that $\langle H \rangle^{Bare}$ is not UV-QD in the $O(4)L\Sigma M$. Appendix 5 reminds the reader (and proves) that UV-QD do not actually arise in

SM 1-oblique-loop HPEW physics and $\langle H \rangle^{Bare}$, which receive at worst only logarithmically divergent corrections. Appendix 6 shows that our 1-loop SM results can (almost certainly) be extended to all QCD and electroweak loop-orders.

Earlier versions of this paper (arXiv [hep-ph]:1106.6354, v1 30 June 2011, v3 27 September 2011) are too long and introduce too many separate, yet important, ideas. The focus here is to show that the SM has no HFTP, rather than explore, explain and dwell on the gauged $O(4)L\Sigma M$ embedded in the SM. Therefore, a separate, more detailed and pedagogical paper has been written [10], which focuses on un-gauged bosonic $O(4)L\Sigma M$ and gives complete and proper explanation and detailed connection to B.W.Lee and K.Symanzik's powerful results [11,12] in that theory and its extension to include virtual SM quarks and leptons. Ref. [10] gives a simple intuitive understanding of the SM results presented here, which arise largely from the embedding of gauged $O(4)L\Sigma M$ into the SM.

**2: $O(4)$ linear sigma model ($O(4)L\Sigma M$) with either massive pions $m_\pi^2 \neq 0$ or Nambu-Goldstone Boson (NGB) pions $m_\pi^2 \to m_{\pi;NGB;L\Sigma M}^2 = 0$**

> *"Whether you like it or not, you have to include in the Lagrangian all possible counter-terms consistent with locality and power counting, unless otherwise constrained by Ward identities"*. Kurt Symanzik (1970) letter to Raymond Stora.

In order to clarify and simplify the classification and disposition of UV-QD arising from the gauged $O(4)L\Sigma M$ embedded in the SM, Section 2 studies the stand-alone un-gauged $O(4)L\Sigma M$ and its UV-QD renormalization. We follow closely (and quote liberally from) the strategy, technology, and pedagogy of the work of Benjamin W. Lee, Kurt Symanzik, N.N.Bogoliubov, O.Parasiuk, K.Hepp, W.Zimmermann (**BPHZ**) and others [11,12]: i.e. the rock upon which modern renormalization theory is built. We also summarize those results of un-gauged $O(4)L\Sigma M$ in Ref. [10] necessary for understanding the classification and disposition of its UV-QD. In Section 3B, we will show that the corresponding gauge invariant subset of 1-loop UV-QD in the SM (i.e. graphs without transverse gauge bosons or ghosts) is equal to the 1-loop UV-QD arising in the un-gauged $O(4)L\Sigma M$ studied here.

**2A: Bosonic $O(4)L\Sigma M$ in the $\langle H \rangle$ vs. $m_\pi^2 / \lambda^2$ half-plane**

We follow closely the pedagogy (and much of the notation) of B.W. Lee [11], and use the linear representation of $\Phi$ to make manifest the physical content of Higgs Vacuum Stability Condition (**VSC**) tadpole renormalization as well as the explicit not-fine-tuned cancellation of UV-QD during Goldstone-mode Symmetry Restoration Condition (**SRC**) enforcement of the Goldstone theorem, i.e. as the pions become NGB and $m_\pi^2 \to m_{\pi;NGB;L\Sigma M}^2 = 0$. We begin with the pure scalar bare Lagrangian, and focus on its UV-QD counter-terms/:

$$L_{L\Sigma M}^{Bare;\Lambda^2} = -|\partial_\mu \Phi|^2 - V_{L\Sigma M}^{Bare;\Lambda^2};$$

$$V_{L\Sigma M}^{Bare;\Lambda^2} = \lambda^2 \left( \Phi^\dagger \Phi - \frac{1}{2}\langle H \rangle^2 \right)^2 - L_{L\Sigma M}^{CounterTerm;\Lambda^2};$$

$$L_{L\Sigma M}^{CounterTerm;\Lambda^2} = L_{L\Sigma M}^{CounterTerm;\Lambda^2}\bigg|_{Symmertric} + L_{L\Sigma M}^{CounterTerm;\Lambda^2}\bigg|_{SymmetryBreaking}$$

$$L_{L\Sigma M}^{CounterTerm;\Lambda^2}\Big|_{Symmertric} = -\delta\mu^2\left[\Phi^\dagger\Phi - \frac{1}{2}\langle H\rangle^2\right];$$

$$L_{L\Sigma M}^{CounterTerm;\Lambda^2}\Big|_{SymmetryBreaking} = \varepsilon_{L\Sigma M}H;$$

$$\lambda^2 > 0; \quad \Phi = \frac{1}{\sqrt{2}}\begin{bmatrix} H + i\pi_3 \\ -\pi_2 + i\pi_1 \end{bmatrix}; \quad \pi_\pm = \frac{1}{\sqrt{2}}(\pi_1 \mp i\pi_2); \tag{2A.1}$$

$$H = h + \langle H\rangle; \quad \langle\Phi\rangle = \frac{\langle H\rangle}{\sqrt{2}}\begin{bmatrix} 1 \\ 0 \end{bmatrix}; \quad \langle h\rangle = 0;$$

where vacuum energy/bubbles are neglected. We assume $\lambda^2 \sim O(1)$ and loosely refer to the real scalar $h$ as the physical "Higgs". Since this paper is only interested in UV-QD, $\delta\mu^2$ has been included entirely in the counter-term in Eq. (2A.1). More traditionally, $V_{L\Sigma M}^{Bare;\Lambda^2}$ is written

$$V_{L\Sigma M}^{Bare;\Lambda^2} = \lambda^2(\Phi^\dagger\Phi)^2 + \mu_{Bare}^2\Phi^\dagger\Phi - L_{L\Sigma M}^{SymmtryBreaking};$$

$$\delta\mu^2 = \mu_{Bare}^2 + \lambda^2\langle H\rangle^2; \tag{2A.2}$$

$$L_{L\Sigma M}^{SymmtryBreaking} = \varepsilon_{L\Sigma M}H;$$

with $\mu_{Bare}^2$ made responsible for absorbing UV-QD. Following [11,12, K.Symanzik above], an explicit symmetry-breaking counter-term $L_{L\Sigma M}^{CounterTerm;\Lambda^2}\Big|_{SymmetryBreaking}$ is included. Apart from that, the theory has $O(4)$ symmetry with conserved vector currents (CVC) $\vec{V}_\mu$ and partially conserved axial-vector currents (PCAC) $\vec{A}_\mu$ where

$$\partial_\mu\vec{V}_\mu = 0;$$
$$\partial_\mu\vec{A}_\mu = -\varepsilon_{L\Sigma M}\vec{\pi}; \tag{2A.3}$$

The first of the connected Green's function Ward-Takahashi identities, connecting the vacuum with the on-shell one-pion state of momentum $q_\mu$, reads [11,12]

$$\langle 0|A_\mu^i(x)|\pi^j(q)\rangle \equiv -i\delta^{ij}\langle H\rangle q_\mu\hat{\pi}^i e^{iq_\mu x_\mu};$$

$$\partial_\mu\langle 0|A_\mu^i(x)|\pi^j(q)\rangle = \delta^{ij}\langle H\rangle q^2\hat{\pi}^i e^{iq_\mu x_\mu} = -\delta^{ij}\langle H\rangle m_\pi^2\hat{\pi}^i e^{iq_\mu x_\mu}; \tag{2A.4}$$

$$\partial_\mu\vec{A}_\mu = -\langle H\rangle m_\pi^2\vec{\pi};$$

so that, including <u>all-orders</u> UV-QD, but ignoring logarithmically divergent and finite contributions:

$$\varepsilon_{L\Sigma M} = \langle H\rangle m_\pi^2;$$

$$L_{L\Sigma M}^{CounterTerm;\Lambda^2}\Big|_{SymmetryBreaking} = \langle H\rangle m_\pi^2 H; \tag{2A.5}$$

where $m_\pi^2$ is the physical renormalized pseudo-scalar pion (pole) mass-squared. $\langle H\rangle^2$ is the experimental value of the renormalized Higgs VEV-squared: i.e. the $O(4)L\Sigma M$ analog of the SM HPEW muon decay constant $\langle H_{SM}\rangle^2 = \left(2\sqrt{2}G_{Muon}^{Experimental}\right)^{-1}$. Lee/Symanzik prove that Eq. (2A.5) receives, to all-loop-orders of perturbation theory, no UV-QD corrections [11,12]:

only logarithmic divergences and finite corrections from wave function renormalization. They then analyze the generic set of $O(4)L\Sigma M$ in the $\langle H \rangle$ vs. $m_\pi^2/\lambda^2$ half-plane.

**2B: Inclusion of 1-loop $O(4)L\Sigma M$ UV-QD**

**B.W.Lee and K.Symanzik proved separately, more than forty years ago, that there are no UV-QD contributions (to all orders of perturbation theory) to $\langle H \rangle^{Bare}$ in $O(4)L\Sigma M$**: it receives only logarithmic divergences [11,12] and finite corrections. Including all UV-QD and logarithmic divergences, K.Symanzik gives a careful BPHZ [12] proof of this fact, with attention to each and every necessary detail: crucial insight is provided by his quotation above. We outline B.W. Lee's 1970 proof that $\langle H \rangle^{Bare}$ is not UV-QD [11] in Appendix 4. Since, in this paper, we are only interested in UV-QD, we will not distinguish between $\langle H \rangle^{Bare}, \langle H \rangle^{Renormalized}, \langle H \rangle^{Experimental}$, calling them all simply $\langle H \rangle$.

We now turn attention to the UV-QD graphs. The UV-QD 1-loop $O(4)L\Sigma M$ Lagrangian, evaluated at zero momentum, including all 1-loop 2-point self-energy and 1-loop 1-point tadpole function UV-QD, is:

$$L_{L\Sigma M}^{1-Loop;\Lambda^2} = C_{L\Sigma M}^{1-Loop;\Lambda^2} \Lambda^2 \left( \Phi^\dagger \Phi - \frac{1}{2}\langle H \rangle^2 \right); \tag{2B.1}$$

with $C_{L\Sigma M}^{1-Loop;\Lambda^2}$ a finite constant [10]. The form of Eq. (2B.1) follows from Lee and Symanzik's proof [11,12] that the theory is properly renormalized, throughout the $\langle H \rangle$ vs. $m_\pi^2/\lambda^2$ plane, with the same UV-QD graphs and counter-terms as in the symmetric "Wigner mode" limit: i.e. $\langle H \rangle^2 \to 0$ holding $m_\pi^2 \neq 0$. Ref. [10] shows that

$$C_{L\Sigma M}^{1-Loop;\Lambda^2} \Lambda^2 = \left(-6\lambda^2\right) \frac{\Lambda^2}{16\pi^2}; \tag{2B.2}$$

and that this result is independent of whether *n*-dimensional or Pauli-Villars regularization is used. To fully exacerbate and reveal any HFTP, we imagine $\Lambda \sim M_{Planck}$ near the Planck scale. Using $L_{L\Sigma M}^{Bare;\Lambda^2}$, we form a 1-loop-UV-QD-improved effective Lagrangian, which includes all scalar 2-point self-energy and 1-point tadpole 1-loop UV-QD, but ignores 1-loop logarithmic divergences, finite contributions and vacuum energy/bubbles:

$$\begin{aligned} L_{L\Sigma M}^{Effective;1-Loop;\Lambda^2} &= L_{L\Sigma M}^{Bare;\Lambda^2} + L_{L\Sigma M}^{1-Loop;\Lambda^2} \\ &= \left|\partial_\mu \Phi\right|^2 - V_{L\Sigma M}^{Renormalized;1-Loop;\Lambda^2}; \end{aligned} \tag{2B.3}$$

$$\begin{aligned} V_{L\Sigma M}^{Renormalized;1-Loop;\Lambda^2} &= \lambda^2 \left[ \frac{1}{2}h^2 + \frac{1}{2}\pi_3^2 + \pi_+\pi_- + \langle H \rangle h \right]^2 \\ &+ m_\pi^2 \left[ \frac{1}{2}h^2 + \frac{1}{2}\pi_3^2 + \pi_+\pi_- \right] + \left( \langle H \rangle m_\pi^2 - \varepsilon_{L\Sigma M} \right)h; \end{aligned} \tag{2B.4}$$

$$m_\pi^2 = \delta\mu^2 - C_{L\Sigma M}^{1-Loop;\Lambda^2} \Lambda^2; \tag{2B.5}$$

is the physical renormalized pseudo-scalar pion (pole) mass-squared in Eqs. (2A.3 to 2A.5).

**Higgs Vacuum Stability Condition (Higgs VSC): The physical Higgs $h$ particle must not simply disappear into the exact UV-QD-corrected vacuum.** Insight into the tadpole term

in Eq. (2B.4) follows by carefully defining the properties of the vacuum, including all effects of UV-QD, and imposing, i.e. on the vacuum and excited states, the Higgs VSC:
1. Includes all perturbative UV-QD corrections, including
    - 1-loop $O(4)L\Sigma M$ when referenced in Sections 2B, 2C, 2D, 2E;
    - 1PI multi-loop $O(4)L\Sigma M$ when so referenced in Section 2D;
    - 1-loop Standard Model when referenced in Section 3C;
    - 1PI multi-loop SM when referenced in Section 3D and Appendix 6;
2. By definition $H = h + \langle H \rangle$ and $\langle (1,0)\Phi \rangle^2 = \frac{1}{2}\langle H \rangle^2$ is the <u>exact</u> VEV. A crucial observation/fact is that, for fixed $m_\pi^2$, as required by the Ward identity, $\langle H \rangle$ minimizes the effective renormalized potential $V_{L\Sigma M}^{\text{Renormalized};1-Loop;\Lambda^2}$ in Eq. (2B.7);
3. $\langle H \rangle$ gets its dimensions from those of the scalar field $\Phi$ [11,12,Appendix 4];
4. The physical Higgs particle $h$ has exactly zero VEV, $\langle h \rangle = 0$;
5. Exact tadpole renormalization is to be imposed to all orders in perturbation theory;
6. It is important to recognize that <u>Higgs VSC tadpole renormalization does not constitute fine-tuning: rather it is a stability condition on the vacuum and excited states of the theory.</u>

Imposition of the Higgs VSC in Eq. (2B.4) requires:
$$\left(\langle H \rangle m_\pi^2 - \varepsilon_{L\Sigma M}\right)h = 0; \tag{2B.6}$$
But the symmetry of the theory, as manifested Eq. (2A.3,2A.4), already insists that $\varepsilon_{L\Sigma M} = \langle H \rangle m_\pi^2$ in Eq. (2A.5). **The Higgs VSC is therefore automatically enforced by the Ward-Takahashi identity [11,12]**. After tadpole renormalization, the effective 1-loop Lagrangian (keeping UV-QD, but ignoring logarithmic divergences, finite parts and vacuum energy/bubbles), can be re-written:

$$L_{L\Sigma M}^{\text{Effective};1-Loop;\Lambda^2} = -\left|\partial_\mu \Phi\right|^2 - V_{L\Sigma M}^{\text{Renormalized};1-Loop;\Lambda^2};$$

$$V_{L\Sigma M}^{\text{Renormalized};1-Loop;\Lambda^2} = \lambda^2 \left[\Phi^\dagger \Phi - \frac{1}{2}\left(\langle H \rangle^2 - \frac{m_\pi^2}{\lambda^2}\right)\right]^2 - \langle H \rangle m_\pi^2 H$$

$$m_\pi^2 = \delta\mu^2 - C_{L\Sigma M}^{1-Loop;\Lambda^2}\Lambda^2; \quad m_h^2 = m_\pi^2 + 2\lambda^2\langle H\rangle^2;$$

$$\Phi^\dagger \Phi = \frac{1}{2}H^2 + \frac{1}{2}\pi_3^2 + \pi_+\pi_-;$$
(2B.7)

**To correctly find $\langle \Phi \rangle$, the Ward-Takahashi identity $\varepsilon_{L\Sigma M} = \langle H \rangle m_\pi^2$ must be enforced while minimizing the renormalized potential.** But such enforcement is automatic in Eq. (2B.7), because $V_{L\Sigma M}^{\text{Renormalized};1-Loop;\Lambda^2}$ is written there in terms of the (fixed) physical renormalized pion pole mass-squared $m_\pi^2$: **the minimum is, of course, at $H = \langle H \rangle$.** Because of the explicit symmetry breaking term $\sim \langle H \rangle m_\pi^2 H$ in $V_{L\Sigma M}^{\text{Renormalized};1-Loop;\Lambda^2}$ in Eq. (2B.7), it is not possible to change the results of Section 2 by going to the unitary $\Phi$ representation [1] or by re-scaling $\langle H \rangle$ (see Appendix 2). We observe that Ward-Takahashi identities, together with Higgs VSC tadpole renormalization, are sufficient to force all 1-loop UV-QD in the un-gauged $O(4)L\Sigma M$ S-Matrix to be absorbed into the physical renormalized pseudo-scalar pion (pole) mass-squared $m_\pi^2$.

## 2C. Wigner mode $m_\pi^2 \neq 0$ and definition of the Higgs Fine-Tuning Problem (HFTP)

For $m_\pi^2 \neq 0$, the 1-loop-UV-QD-improved effective $O(4)L\Sigma M$ Lagrangian has 1-loop UV-QD contributions from $L_{L\Sigma M}^{1-Loop;\Lambda^2}$ to both the pion and Higgs masses. These are cancelled by the counter-term $L_{L\Sigma M}^{CounterTerm;\Lambda^2}\big|_{Symmertric}$, but may leave very large finite residual contributions:

$$m_\pi^2 = \delta\mu^2 + (6\lambda^2)\frac{\Lambda^2}{16\pi^2};$$
$$m_h^2 = 2\lambda^2\langle H\rangle^2 + \delta\mu^2 + (6\lambda^2)\frac{\Lambda^2}{16\pi^2} = 2\lambda^2\langle H\rangle^2 + m_\pi^2;$$
(2C.1)

As expected, the "natural" scale of the coefficient of the relevant symmetric scalar 2-point operator is $\delta\mu^2 \sim \Lambda^2$. Now imagine that weak interaction experiments require $G_{Muon}^{Experimwental}\langle H\rangle^2 \sim O(1)$ with $G_{Muon}^{Experimwental} = 1.16637 \times 10^{-5} GeV^{-2} = (292.807 GeV)^{-2}$ the weak scale, set by the experimental muon decay constant. The Higgs VEV $\langle H\rangle$ is a free parameter, shown to receive no UV-QD corrections: only at worst logarithmically divergent corrections [11,12,Appendix 4]. It is not otherwise determined by the internal self-consistency of the theory, so there is no problem in determining its numerical value from experiment. But problems appear if experiments require weak-scale $G_{Muon}^{Experimwental} m_h^2, G_{Muon}^{Experimwental} m_\pi^2 \sim O(1)$. For Planck scale ultra-violet cut-off $\Lambda \sim M_{Planck}$, absorption of the UV-QD into $m_h^2, m_\pi^2$ requires fine-tuning $\delta\langle H\rangle^2$ to within $(G_{Muon}^{Experimwental} M_{Planck}^2)^{-1} \sim 10^{-32}$ of its natural value [2,13,14], a HFTP and violation of "Naturalness" which has been variously defined as the demand that
- Observable properties of the theory be stable against minute variations of fundamental parameters [14];
- Electroweak radiative corrections be the same order (or much smaller) than the actually observed values [2];

$O(4)L\Sigma M$ in "Wigner mode", with unbroken $O(4)$ symmetry and four degenerate massive scalars (i.e. 1 scalar + 3 pseudo-scalars), is defined as the limit $\varepsilon_{L\Sigma M} = \langle H\rangle m_\pi^2 \to 0$, holding $m_\pi^2 \neq 0$ while $\langle H\rangle \to 0$ [11,12]. Including 1-loop UV-QD (but ignoring logarithmic divergences, finite contributions and vacuum energy/bubbles) the 1-loop-corrected effective Lagrangian is:

$$L_{L\Sigma M}^{Effective;1-Loop;\Lambda^2} \xrightarrow[m_\pi^2 \neq 0]{\langle H\rangle \to 0} L_{WignerMode;L\Sigma M}^{Effective;1-Loop;\Lambda^2};$$
(2C.2)

$$L_{WignerMode;L\Sigma M}^{Effective;1-Loop;\Lambda^2} = -|\partial_\mu \Phi|^2 - \lambda^2\left[\Phi^\dagger\Phi + \frac{1}{2}\frac{m_\pi^2}{\lambda^2}\right]^2;$$

$$H \to h; \quad \langle H \rangle \to 0;$$

$$m_h^2 = m_\pi^2 = \delta\mu^2 + \left(6\lambda^2\right)\frac{\Lambda^2}{16\pi^2}; \tag{2C.3}$$

If the Higgs and pion masses are fine-tuned to an un-natural $m_h^2, m_\pi^2 \ll \Lambda^2$ "experimental" value (e.g. the weak scale), Wigner mode $O(4)L\Sigma M$ suffers the classic HFTP, currently regarded as fatal mathematical flaws in all $O(4)L\Sigma M$ and the SM. *"…scalar fields…are not protected from acquiring large bare masses by any symmetry of the SM, so it is difficult to see why their masses, and hence all other masses, are not in the neighbourhood of $10^{16}$ to $10^{18}$ GeV* [S.Weinberg in Ref. 22]. In contrast, we show in sections 2D and 3C respectively, that the Higgs in Goldstone-mode $O(4)L\Sigma M$ and the SM is protected from acquiring a large bare mass by the symmetries of those theories as realized by the Goldstone theorem after spontaneous symmetry breaking (SSB). SSB opens a loop-hole in the logic of the HFTP!

**2D. Goldstone mode weak scale spontaneously broken $O(4)L\Sigma M$ (i.e. with $m_\pi^2 \to m_{\pi;NGB;L\Sigma M}^2 = 0$ identically and exactly) has zero UV-QD remnant and does not suffer a HFTP**

In contrast, Goldstone-mode-$O(4)L\Sigma M$, with spontaneously broken $O(4)$ symmetry, 1 massive scalar and 3 exactly mass-less pseudo-scalars, is defined as the opposite limit $\varepsilon_{L\Sigma M} = \langle H \rangle m_\pi^2 \to 0$, holding $\langle H \rangle \neq 0$ while $m_\pi^2 \to 0$ [11,12]. Including 1-loop UV-QD, but ignoring logarithmic divergences, finite contributions and vacuum energy/bubbles, the 1-loop-corrected effective Lagrangian is

$$L_{L\Sigma M}^{Effective;1-Loop;\Lambda^2} \xrightarrow[m_\pi^2 \to 0]{\langle H \rangle \neq 0} L_{GoldstoneModeL\Sigma M}^{Effective;1-Loop;\Lambda^2}; \tag{2D.1}$$

$$L_{GoldstoneModeL\Sigma M}^{Effective;1-Loop;\Lambda^2} = -\left|\partial_\mu \Phi\right|^2 - \lambda^2 \left[\Phi^\dagger \Phi - \frac{1}{2}\langle H \rangle^2\right]^2;$$

$$H = h + \langle H \rangle; \tag{2D.2}$$

$$m_\pi^2 = \delta\mu^2 - C_{L\Sigma M}^{1-Loop;\Lambda^2}\Lambda^2 \to m_{\pi;NGB;L\Sigma M}^2 = 0;$$

$$m_h^2 \to 2\lambda^2 \langle H \rangle^2;$$

Insight into these equations follows again, by carefully defining the properties of the vacuum, including all effects of 1-loop-induced UV-QD, after SSB:

**Goldstone Symmetry Restoration Condition (Goldstone SRC): After SSB, the Goldstone theorem must be an exact property (to all loop orders) of the Goldstone-mode $O(4)L\Sigma M$ vacuum and its excited states.**
1. Includes all perturbative UV-QD corrections, including
    - 1-loop $O(4)L\Sigma M$ when referenced in Section 2D, 2E;
    - 1PI Multi-loop $O(4)L\Sigma M$ when referenced in Section 2D;
    - 1-loop SM when referenced in Section 3C;
    - 1PI Multi-loop SM when referenced in Section 3D and Appendix 6;
2. In generic $O(4)L\Sigma M$ theories, spanning the $\langle H \rangle$ vs. $m_\pi^2/\lambda^2$ half-plane, it is necessary to impose this Goldstone-SRC (essentially by hand) in order to force the theory to the

Goldstone-mode-$O(4)L\Sigma M$ limit, i.e. onto the line $m_\pi^2 \to m_{\pi;NGB;L\Sigma M}^2 = 0$. $m_{\pi;NGB;L\Sigma M}^2$ is therefore a loop-induced NGB mass-squared (see Ref. [10] and Appendix 2);

3. In more modern language, it is necessary to explicitly <u>enforce</u> the Goldstone theorem so that $\pi_+, \pi_-, \pi_3$ remain exactly mass-less to all orders of perturbation theory. That is the purpose of Lee/Symanzik's Goldstone-SRC;
4. Since UV-QD (in the generic $O(4)L\Sigma M$ S-Matrix) are all packed into the physical renormalized pseudo-scalar pion (pole) mass-squared, the Goldstone theorem also forces any finite remnant of UV-QD to be exactly zero in the Goldstone mode $m_\pi^2 \to m_{\pi;NGB;L\Sigma M}^2 = 0$ limit.
5. We have ignored certain infra-red (**IR**) subtleties as beyond the scope of this paper.

It is easy to see that $V_{L\Sigma M}^{\text{Renormalized};1-loop;\Lambda^2}$ in Eq. (2B.7) only has NGB when "bottom of the wine bottle Goldstone symmetry" is restored: i.e. in Goldstone mode with

$$m_\pi^2 \to m_{\pi;NGB;L\Sigma M}^2 = 0 \quad (2D.3)$$

The crucial observation about Eqs. (2D.1,2D.2) is that <u>proper 1-loop enforcement of the Goldstone theorem requires imposition (essentially by hand) of Lee/Symanzik's Goldstone-SRC [11,12]. The Goldstone theorem then forces</u>

$$0 = m_{\pi;NGB;L\Sigma M}^2 \left(\frac{1}{2}h^2 + \frac{1}{2}\pi_3^2 + \pi_+\pi_-\right);$$

$$0 = m_{\pi;NGB;L\Sigma M}^2 = \delta\mu^2 + \left(6\lambda^2\right)\frac{\Lambda^2}{16\pi^2}; \quad (2D.4)$$

<u>identically, with exactly zero UV-QD remnant.</u>

A central observation here and in Ref. [10] is that all finite remnants of 1-loop UV-QD contributions to the Higgs' mass in Eq. (2D.4) to vanish identically, without fine-tuning:

$$\left(\delta\mu^2 - C_{L\Sigma M}^{1-Loop;\Lambda^2}\Lambda^2\right)\left(\frac{1}{2}h^2\right) = m_{\pi;NGB;L\Sigma M}^2\left(\frac{1}{2}h^2\right) = 0; \quad (2D.5)$$

$$m_h^2 = 2\lambda^2 \langle H \rangle^2; \quad (2D.6)$$

Lee/Symanzik proved that $\langle H_{Bare} \rangle^2$ is not UV-QD [11,12] and receives at worst logarithmic divergences. B.W.Lee's 1970 proof [11] of this fact is outlined in Appendix 4. Therefore, **the renormalized Higgs mass-squared in Eq. (2D.6) is not fine-tuned and Goldstone mode $O(4)L\Sigma M$ has no 1-loop HFTP!**

Most of the calculations in Section 2 are not new: rather, they have been common $O(4)L\Sigma M$ knowledge for more than four decades. What is new here and in Ref. [10] are observations that, in <u>spontaneously broken</u> $O(4)L\Sigma M$ (i.e. in Goldstone mode):

1. Although the coefficients of the relevant dimension-2 scalar self-energy-2-point and tadpole-1-point relevant operators take their "natural" scale $\delta\mu^2 \sim \Lambda^2$ (e.g. even with the UV cut-off taken as Planck scale $\Lambda \sim M_{Planck}$), there is no need to fine-tune a "weak-scale Higgs" mass $m_h^2 G_{Muon}^{Experimental} \sim O(1)$ if experiment so demands;
2. Our no-fine-tuning-theorem for a weak-scale Higgs mass is then simply another (albeit un-familiar) consequence of the Goldstone theorem, an exact property of the spontaneously broken $O(4)L\Sigma M$ vacuum and spectrum;

3. <u>Lee/Symanzik's Lesson from this Section 2D</u>: Spontaneously broken $O(4)L\Sigma M$ must be viewed as a limiting case of Lee/Symanzik's generic set of $m_\pi^2 \geq 0$ theories (i.e. the $m_\pi^2 = 0$ line in the $\langle H \rangle$ vs. $m_\pi^2/\lambda^2$ plane) where two conditions must be imposed (essentially by hand) on the Goldstone mode $O(4)L\Sigma M$ vacuum and excited states
   - Higgs SC: the Higgs must not simply disappear into the vacuum;
   - Goldstone-SRC: masses of NGB must be exactly zero;

An important subtlety arising in Goldstone-mode-$O(4)L\Sigma M$, with strong analogy in the SM, is discussed in Appendix 2. If the calculations and analysis of Section 2D are re-done, but <u>beginning</u> with the Goldstone-mode-$O(4)L\Sigma M$ bare Lagrangian (i.e. Eq. (2A.1) with $\varepsilon_{L\Sigma M} = 0$)

$$L^{Bare}_{GoldstoneModeL\Sigma M} = -\left|\partial_\mu \Phi\right|^2 - \lambda^2 \left(\Phi^\dagger \Phi - \frac{1}{2}\langle H \rangle^2\right)^2 - \delta\mu^2 \left(\Phi^\dagger \Phi - \frac{1}{2}\langle H \rangle^2\right); \quad (2D.7)$$

one must still prescribe and enforce $m_\pi^2 \to m_{\pi;NGB;L\Sigma M}^2 = 0$. The self-consistent "Goldstone Mode UV-QD Renormalization Prescription" (GMRP) is required to avoid a "fine-tuning discontinuity". Because the <u>tree-level</u> Standard Model is "already in Goldstone mode", Appendix 2 gives necessary insight into the imposition of Lee/Symanzik's two renormalization conditions for the case of the Standard Model in Sections 3C and 3D.

***Section 2E and Appendix D of Ref. [10] show that Goldstone mode weak scale spontaneously broken $O(4)L\Sigma M$ (i.e. with $m_\pi^2 \to m_{\pi;NGB;L\Sigma M}^2 = 0$ identically and exactly) has zero remnant of UV-QD and does not suffer HFTP to any perturbative loop-order.***

**2E: UV-QD in Goldstone mode $O(4)L\Sigma M$ with SM quarks and leptons [10]**

All-orders renormalization of the generic set of bosonic $O(4)L\Sigma M$ was long ago extended to include nucleons by J.L Gervais & B.W.Lee and K.Symanzik in Ref. [12], and is easily instead extended to include SM quarks and leptons. The total UV-QD 1-loop Lagrangian, including the effects of SM quarks and leptons, is re-calculated in Ref. [10]

$$L^{1-Loop;\Lambda^2}_{L\Sigma M+SMQuarks\&Leptons} = C^{1-Loop;\Lambda^2}_{L\Sigma M+SMQuarks\&Leptons} \Lambda^2 \left(\Phi^\dagger \Phi - \frac{1}{2}\langle H \rangle^2\right); \quad (2E.1)$$

$$C^{1-Loop;\Lambda^2}_{L\Sigma M+SMQuarks\&Leptons} \Lambda^2 = \left(\frac{-3m_h^2 + 4\sum_{Quarks}^{Flavor,Color} m_{Quark}^2 + 4\sum_{Leptons}^{Flavor} m_{Lepton}^2}{16\pi^2 \langle H \rangle^2}\right)\Lambda^2; \quad (2E.2)$$

and is recognizable as the zero-gauge-coupling limit of 1-loop UV-QD in the SM [2]. Ref. [10] shows that UV-QD remnants from virtual SM quarks and leptons vanish identically, and that there is no HFTP to all perturbative loop orders in Goldstone mode $O(4)L\Sigma M$. Ignoring logarithmic divergences, finite contributions and vacuum energy/bubbles, Ref. [10] displays the scalar-sector 1-loop-UV-QD-improved <u>effective Goldstone-mode Lagrangian</u>, including 1-loop 2-point self-energies and 1-loop 1-point tadpole UV-QD from virtual scalars and SM quarks and leptons, after imposition of Lee/Symanzik's two UV-QD renormalization conditions:

$$L^{Effective;;1-Loop;\Lambda^2}_{GoldstoneMode;L\Sigma M+SMQuarks\&Leptons;\Phi-Sector} = -|\partial_\mu \Phi|^2 - \lambda^2 \left(\Phi^\dagger \Phi - \frac{1}{2}\langle H \rangle^2\right)^2; \quad (2E.3)$$

$$m^2_{\pi;NGB;L\Sigma M+SMQuarks\&Leptons} = \delta\mu^2 - C^{1-Loop;\Lambda^2}_{L\Sigma M+SMQuarks\&Leptons}\Lambda^2 = 0; \quad (2E.4)$$

$$m^2_h = 2\lambda^2\langle H \rangle^2; \quad H = h + \langle H \rangle; \quad \langle h \rangle = 0; \quad (2E.5)$$

It is easy to extend B.W.Lee's all-orders proof [11,Appendix 4], i.e. that $\langle H \rangle^2$ and $m_h^2$ in Eq. (2E.5) are not UV-QD, receive at worst logarithmic divergences, and are not fine tuned: Eqs. (2E.3-5) generalize to all-loop-orders, $L^{Effective;;All-Loop;\Lambda^2}_{GoldstoneMode;L\Sigma M+SMQuarks\&Leptons;\Phi-Sector}$ does not depend on $C^{All-Loop;\Lambda^2}_{L\Sigma M+SMQuarks\&Leptons}$ and **Goldstone mode $O(4)L\Sigma M$ with SM quarks and leptons has no HFTP to any order of loop perturbation theory.**

### 3: Standard $SU(3)_{Color} \times SU(2)_{Isospin} \times U(1)_{Hypercharge}$ Model

The stand-alone SM particle/field content is:
- Color, isospin and hypercharge gauge bosons: $G_\mu^A, W_\mu^a, B_\mu$;
- Linear representation of $O(4)L\Sigma M$ scalars: $\Phi = \frac{1}{\sqrt{2}}\begin{bmatrix} H + i\pi_3 \\ -\pi_2 + i\pi_1 \end{bmatrix}$;
- Color, isospin and hypercharge ghosts: $\bar{\xi}^A, \eta^A; \bar{\xi}^a, \eta^a; \bar{\xi}, \eta$;
- Indices: $Lorentz: \mu = 1,4; \quad Color: A = 1,8; \quad Isospin: a = 1,3$;
- Color, isospin and hypercharge gauge couplings: $g_3, g_2, g_1$;
- Fermions: 3 generations of quarks and leptons;
- VEV: $H = h + \langle H \rangle; \quad \Phi = \frac{1}{\sqrt{2}}\begin{bmatrix} \langle H \rangle \\ 0 \end{bmatrix}$;
- Various parameters: CKM matrices, quark and lepton masses, etc;

**3A: 1-loop UV-QD do not arise in 4-massless-external-fermion SM high precision electroweak (HPEW) processes, carried by massive $W_\mu^\pm, Z_\mu$ bosons and photons**

The weak mixing angle, gauge boson masses and Einhorn-Jones-Veltman (EJV) custodial SU2 breaking $\rho_{EJV}$ parameter [15] are related:

$$s_\theta = \sin\theta_{Weak}; \quad c_\theta = \cos\theta_{Weak}; \quad e = g_2 s_\theta = g_1 c_\theta;$$

$$\begin{bmatrix} Z_\mu \\ A_\mu \end{bmatrix} = \begin{bmatrix} c_\theta & -s_\theta \\ s_\theta & c_\theta \end{bmatrix}\begin{bmatrix} W_\mu^3 \\ B_\mu \end{bmatrix}; \quad M_W = \frac{g_2\langle H\rangle}{2}; \quad M_Z^2 = \frac{M_W^2}{c_\theta^2 \rho_{EJV}}; \quad (3A.1)$$

Before analyzing UV-QD in the scalar-sector, it is important to first address possible 1-loop SM UV-QD <u>outside</u> the $\Phi$ sector effective Lagrangian. As the reader is reminded in Appendices 1 and 5, it was proved long ago that **UV-QD do not arise in 1-loop SM massless 4-fermion HPEW processes mediated by massive $W_\mu^\pm, Z_\mu$ bosons and photons [5]: i.e. bare or renormalized SM HPEW parameters are not fine-tuned.**

Most of the important 1-loop SM HPEW effects are embedded in 2-point "oblique" corrections. Gauge invariance requires careful inclusion of universal longitudinal components of weak massive gauge bosons with combinations of oblique loops [5,16,1, Appendix 5]. As a by-product, all remnant low-energy "non-decoupling" 1-oblique-loop effects, of new very heavy Beyond the Standard Model (**BSM**) particles $M_{BSM}^2 >> q^2$, were classified completely into three operators in a HPEW effective SM Lagrangian [5]: the $\rho_{EJV}$ parameter [15] and two new **dimension-zero** gauge invariant functions [5], $\Delta_3/q^2$ and $\Delta_+/q^2$ (see Eqs. (A5.5-8)), which become constants as $q^2/M_{BSM}^2 \to 0$ [5,1,17]. These three non-decoupling parameters $\rho_{EJV}$, $\left[\Delta_3/q^2\right]_{q^2=0}$ and $\left[\Delta_+/q^2\right]_{q^2=0}$ were used to classify all non-decoupling 1-oblique-loop HPEW effects of heavy SM (e.g. top quark and Higgs) as well as BSM particles with SM quantum numbers [5]. Our three parameters were much later popularized by many groups, are now usually called respectively $T, S, U$ [1,17], and have indirectly severely constrained heavy BSM physics [5,17,18], by comparing HPEW theory vs. experiments at low $q^2/M_{BSM}^2$. The most powerful oblique-loop BSM constraints are, as predicted [5,8,18], from LEP1/SLC and $M_W, M_Z$ data.

For this paper, the important point about $\rho_{EJV}, \Delta_3, \Delta_+$ is that each dimension-2 coefficient $M_W^2, M_Z^2, \langle H \rangle^2, G_{Muon}^{-1}$ of relevant SM dimension-2 gauge invariant operators neither receives nor absorbs 1-loop UV-QD corrections [5,1,Appendix 5]: e.g. the $q^2$-dependent finite "running muon decay constant" in 4-massless-fermion high precision HPEW processes [5]

$$\frac{1}{4\sqrt{2}G_{Muon}^*(q^2)} = \frac{1}{2}\langle H_{Bare}\rangle^2 - \text{Re}\left[\Pi_{+-}(q^2) - \Pi_{3Q}^{Transverse}(q^2) - 2\Pi_{3Q}^{Longitudinal}(q^2)\right];$$

$$G_{Muon}^*(0) = G_{Muon}^{Experimental} + CertainVertex \& BoxParts[5];$$
(3A.2)

is written as a gauge invariant [5,16,1,Appendix 5] combination of vector-boson 1-loop 2-point self-energies (e.g. $\Pi_{+-}, \Pi_{3Q}^{Transverse}$), and universal longitudinal parts $\Pi_{3Q}^{Longitudinal}$ (or, almost equivalently, certain universal 3-point vertex and 4-point pinched box functions [16]). Appendix 5 proves that **UV-QD do not arise in 1-loop HPEW processes** [5,1]. Therefore, SM relations such as

$$M_W^2 = M_Z^2 c_\theta^2 \rho_{EJV} = \frac{e^2}{s_\theta^2}\frac{1}{4\sqrt{2}G_{Muon}} = \frac{g_2^2\langle H\rangle^2}{4};$$

$$\langle \Phi_{Bare}\rangle^2 = \frac{1}{2}\langle H\rangle^2 = \frac{1}{2\sqrt{2}G_{Muon}};$$
(3A.3)

receive at worst logarimically divergent corrections at 1-loop.

$\langle\Phi_{Bare}\rangle^2$ **is not a free parameter in the SM scalar effective Lagrangian** (i.e. it has already been determined by the HPEW experimental value of $G_{Muon}^{Experimental}$ [5,1,18]) and **cannot therefore be used to absorb 1-loop UV-QD** from SM scalar-sector 2-point functions: Instead, UV-QD must vanish identically as in Appendix 5, leaving the 1-loop SM relation

$$m_h^2/M_W^2 = 8\lambda^2/g_2^2$$
(3A.4)

with at worst logarithmically divergent corrections. The 1-loop renormalized Higgs VEV $\langle H \rangle$ and its running $\langle H_*(q^2) \rangle$ are derived quantities. Although either equation in (A5.5) can be used to define them, the usual choice is:

$$\langle H_*(q^2) \rangle^2 \equiv \frac{1}{2\sqrt{2}G_\mu^*(q^2)}; \qquad \langle H \rangle^2 \equiv \langle H_*(0) \rangle^2 = \frac{1}{2\sqrt{2}G_\mu^*(0)} = (246 GeV)^{-2}; \qquad (3A.5)$$

$$\langle H \rangle^2 = \langle H_{Bare} \rangle^2 - 2\mathrm{Re}\left[\Pi_{+-}(0) - 2\Pi_{3Q}^L(0)\right]; \qquad (3A.6)$$

**Eq. (3A.6) and Appendix 5 prove that, at 1-loop, the SM bare or renormalized Higgs VEV absorbs at worst logarithmic divergences. UV-QD do not arise in HPEW physics [5] and $\langle H \rangle$ is not fine-tuned!**

**3B: Inclusion of 1-loop UV-QD in the scalar-sector SM effective Lagrangian**

Since the SM is (almost) the $SU(3)_{Color} \times SU(2)_{Isospin} \times U(1)_{Hyperch\,arg\,e}$ gauged version of Goldstone-mode-$O(4)L\Sigma M$ (i.e. with carefully chosen fermion quantum numbers, GIM, etc.), its 1-loop scalar-sector effective Lagrangian receives 1-loop UV-QD corrections analogous with those studied in Section 2 and Ref. [10]. This is (of course) because a gauged version of the $O(4)L\Sigma M$ is embedded in the SM. A sub-set of SM UV-QD arising from that embedded $O(4)L\Sigma M$ (re-calculated in a general $R_\xi$ gauge in Appendix 3) therefore corresponds to the set of UV-QD arising in the stand-alone un-gauged $O(4)L\Sigma M$: specifically, in that corresponding SM sub-set
- SM UV-QD arising from graphs containing at least one virtual transverse gauge boson or ghost are excluded;
- SM UV-QD arising from graphs containing only virtual NGB, Higgs and SM quarks and leptons are included;

All 1-loop UV-QD appearing in the SM were calculated and agreed long ago [2,6]. We want the gauge invariant set of UV-QD arising from 1-loop SM scalar self-energy and tadpole graphs with no virtual gauge bosons or ghosts. As shown in Appendix 3, these contribute terms to the effective SM Lagrangian:

$$L_{SM}^{1-Loop;\Lambda^2}\bigg|_{\substack{NoVirtual \\ GaugesOrGhosts}} = C_{SM}^{1-Loop;\Lambda^2}\bigg|_{\substack{NoVirtual \\ GaugesOrGhosts}} \Lambda^2 \left(\frac{1}{2}h^2 + \frac{1}{2}\pi_3^2 + \pi_+\pi_- + \langle H \rangle h\right); \qquad (3B.1)$$

$$C_{SM}^{1-Loop;\Lambda^2}\bigg|_{\substack{NoVirtual \\ GaugesOrGhosts}} = \left(\frac{-6\lambda^2 \langle H \rangle^2 + 4\sum_{Quarks}^{Flavor,Color} m_{Quark}^2 + 4\sum_{Leptons}^{Flavor} m_{Lepton}^2}{16\pi^2 \langle H \rangle^2}\right); \qquad (3B.2)$$

Careful comparison reveals, not just correspondence, but equality!

$$L_{SM}^{1-Loop;\Lambda^2}\bigg|_{\substack{NoVirtual \\ GaugesOrGhosts}} = L_{L\Sigma M+SMQuarks\&Leptons}^{1-Loop;\Lambda^2}; \qquad (3B.3)$$

The additional UV-QD 1-loop scalar 2-point self-energies and 1-point tadpoles involve virtual gauge bosons and ghosts and vanish smoothly in the zero-gauge-coupling limit. They were calculated and agreed in $R_\xi : \xi = 1$ gauge long ago [2,6]. Appendix 3 re-calculates the appropriate 1-loop self-energy and tadpole graphs in a general $R_\xi$ gauge

$$L_{SM}^{1-Loop;\Lambda^2} = C_{SM}^{1-Loop;\Lambda^2} \Lambda^2 \left( \frac{1}{2} h^2 + \frac{1}{2} \pi_3^2 + \pi_+ \pi_- + C_{SM;\xi}^{1-Loop;\Lambda^2} \langle H \rangle h \right); \tag{3B.4}$$

$$C_{SM}^{1-Loop;\Lambda^2} = \left( \frac{(1-n)(M_Z^2 + 2M_W^2) - 3m_h^2 + 4 \sum_{Quarks}^{Flavor,Color} m_{Quark}^2 + 4 \sum_{Leptons}^{Flavor} m_{Lepton}^2}{16\pi^2 \langle H \rangle^2} \right); \tag{3B.5}$$

with two constants: gauge invariant and famous $C_{SM}^{1-Loop;\Lambda^2}$ [2] and gauge-dependent $C_{SM}^{\xi}$ in Eq. (A3.23). The reader is reminded that Eqs. (3B.1-8) ignore logarithmic divergences and finite contributions.

Focusing on the UV-QD counter-term, the bare scalar sector SM Lagrangian is

$$L_{SM;\Phi-Sector}^{Bare;\Lambda^2} = -|D_\mu \Phi|^2 - \lambda^2 \left[ \Phi^\dagger \Phi - \frac{1}{2} \langle H \rangle^2 \right]^2 + L_{SM;\Phi-Sector}^{CounterTerm;\Lambda^2};$$

$$L_{SM;\Phi-Sector}^{CounterTerm;\Lambda^2} = L_{SM;\Phi-Sector}^{CounterTerm;\Lambda^2} \bigg|_{Symmertric} + L_{SM;\Phi-Sector}^{CounterTerm;\Lambda^2} \bigg|_{GaugeDependent} \tag{3B.9}$$

$$L_{SM;\Phi-Sector}^{CounterTerm;\Lambda^2} \bigg|_{Symmertric} = -\delta\mu^2 \left[ \Phi^\dagger \Phi - \frac{1}{2} \langle H \rangle^2 \right];$$

$$L_{SM;\Phi-Sector}^{CounterTerm;\Lambda^2} \bigg|_{GaugeDependent} = \varepsilon_{SM}^{\xi} H;$$

where, following K.Symanzik (see quote at beginning of Section 2), we include a gauge-dependent counter-term in analogy Eq. (2A.1). We then form a 1-loop-improved scalar-sector effective SM Lagrangian, which includes all SM self-energy 2-point and tadpole 1-point UV-QD (but ignores logarithmically divergent and finite contributions and vacuum energy/bubbles):

$$L_{SM;\Phi-Sector}^{Effective;1-Loop;\Lambda^2} = -|D_\mu \Phi|^2 - V_{L\Sigma M}^{Renormalized;1-Loop;\Lambda^2};$$

$$V_{SM}^{Renormalized;1-Loop;\Lambda^2} = \lambda^2 \left[ \frac{1}{2} h^2 + \frac{1}{2} \pi_3^2 + \pi_+ \pi_- + \langle H \rangle h \right]^2$$

$$+ \left( \delta\mu^2 - C_{SM}^{1-Loop;\Lambda^2} \Lambda^2 \right) \left[ \frac{1}{2} h^2 + \frac{1}{2} \pi_3^2 + \pi_+ \pi_- \right] + \left[ \left( \delta\mu^2 - C_{SM;\xi}^{1-Loop;\Lambda^2} C_{SM}^{1-Loop;\Lambda^2} \Lambda^2 \right) \langle H \rangle - \varepsilon_{SM}^{\xi} \right] h; \tag{3B.9}$$

$$H = h + \langle H \rangle; \quad \langle h \rangle = 0;$$

Imposing Higgs VSC tadpole renormalization, we insist that

$$\left[ \left( \delta\mu^2 - C_{SM;\xi}^{1-Loop;\Lambda^2} C_{SM}^{1-Loop;\Lambda^2} \Lambda^2 \right) \langle H \rangle - \varepsilon_{SM}^{\xi} \right] = 0; \tag{3B.10}$$

Remember that tadpole renormalization does not constitute fine-tuning, but rather is a stability condition on the vacuum and excited states of the theory. Note that, in $R_\xi; \xi = 1$ gauge where $C_{SM;\xi=1}^{1-Loop;\Lambda^2} = 1$, Eq. (3B.10) can be written

$$\varepsilon_{SM}^{\xi=1} = \langle H \rangle \left( \delta\mu^2 - C_{SM}^{1-Loop;\Lambda^2} \Lambda^2 \right) = \langle H \rangle m_{\pi;SM}^2 \tag{3B.11}$$

in analogy with the Ward-Takahashi identity Eq. (2A.5) in un-gauged $O(4)L\Sigma M$.

After tadpole renormalization, the effective 1-loop SM scalar-sector Lagrangian (i.e. keeping UV-QD, but ignoring logarithmic divergences, finite parts and vacuum energy/bubbles) is:

$$L_{SM;\Phi-Sector}^{Effective;1-Loop;\Lambda^2} = -\left|D_\mu \Phi\right|^2 - V_{SM}^{Renormalized;1-Loop;\Lambda^2};$$

$$V_{SM}^{Renormalized;1-Loop;\Lambda^2} = \lambda^2 \left[\Phi^\dagger \Phi - \frac{1}{2}\left(\langle H \rangle^2 - \frac{m_{\pi;SM}^2}{\lambda^2}\right)\right]^2 - \langle H \rangle m_{\pi;SM}^2 H; \qquad (3B.12)$$

$$m_{\pi;SM}^2 = \delta\mu^2 - C_{SM}^{1-Loop;\Lambda^2}\Lambda^2; \qquad m_h^2 = m_{\pi;SM}^2 + 2\lambda^2 \langle H \rangle^2;$$

$$\Phi^\dagger \Phi = \frac{1}{2}H^2 + \frac{1}{2}\pi_3^2 + \pi_+\pi_-; \qquad (3B.13)$$

where the parameters $\lambda^2, \langle H \rangle, m_{\pi;SM}^2$ are gauge invariant. As required, the minimum of the renormalized SM potential Eq. (3B.12) is at $H = \langle H \rangle$. Because of the appearance of the explicit **gauge-symmetry breaking term** $\sim \langle H \rangle m_{\pi;SM}^2 H$ in $V_{SM}^{Renormalized;1-Loop;\Lambda^2}$ in Eq. (3B.12), it is not possible to change the SM results of Section 3 by going to the unitary $\Phi$ representation [1] or by re-scaling $\langle H \rangle$ (see Appendix 2). We observe that Higgs VSC is sufficient to force all 1-loop SM S-Matrix UV-QD to be absorbed into $m_{\pi;SM}^2$. A hint as to the eventual disposition of the gauge-symmetry-breaking term is the observation that $V_{SM}^{Renormalized;1-Loop;\Lambda^2}$ in Eq. (3B.12) only has NGB when "bottom of the wine bottle Goldstone symmetry" is restored: i.e. as $m_{\pi;SM}^2 \to m_{\pi;NGB;SM}^2 = 0$ in the spontaneously broken SM.

**3C: At 1-loop, the SM has no surviving remnant of UV-QD and does not suffer a HFTP**

We have established a correspondence between the UV-QD renormalizations of the SM and the un-gauged $O(4)L\Sigma M$ with SM quarks & leptons. Since

- $C_{L\Sigma M+SMQuarks\&Leptons}^{1-Loop;\Lambda^2}$ in Eq. (2E.2) is the smooth zero-gauge-coupling limit of $C_{SM}^{1-Loop;\Lambda^2}$ in Eq. (3B.8), as proved in Eq. (3B.2) and Appendix 3;
- The contribution $L_{L\Sigma M+SMQuarks\&Leptons}^{1-Loop;\Lambda^2}$ to the effective Lagrangian in Eq. (2E.1) is the smooth zero-gauge-coupling limit of $L_{SM}^{1-Loop;\Lambda^2}$ in Eq. (3B.4), as proved in Eq. (3B.1) and Appendix 3;

we insist that self-consistent renormalization of the SM requires that

- These pairs of 1-loop UV expressions be treated the same way;
- The scalar-sector effective Lagrangians of the SM, Eq. (3B.12), and un-gauged Goldstone-mode $O(4)L\Sigma M$ with SM quarks & leptons, Eqs. (2E.3, 2E.4), be identical

$$L_{SM;\Phi-Sector}^{Effective;1-Loop;\Lambda^2} \xrightarrow[g_1,g_2,g_3 \to 0]{} L_{GoldstoneMode;L\Sigma M+SMQuarks\&Leptons;\Phi-Sector}^{Effective;1-Loop;\Lambda^2} \qquad (3C.1)$$

in the appropriate zero-gauge coupling limit. The limit must be smooth: i.e. without a "fine tuning discontinuity", as defined in Appendix 2. Eq. (3C.1) is enforced, including all 1-loop UV-QD (but ignoring logarithmic divergences, finite contributions and vacuum energy/bubbles), by interpreting Lee/Symanzik's lesson on un-gauged $O(4)L\Sigma M$ at the end of Section 2D carefully:

**Self-consistent renormalization of the SM carries instruction to impose (essentially by hand) Lee/Symanzik's two conditions on the resultant SM vacuum and excited states: i.e. the SM Goldstone mode Renormalization Prescription (GMRP):**
- Higgs VSC: the Higgs must not simply disappear into the vacuum;

- Goldstone SRC: the loop-induced mass-squared of SM NGB must be identically zero, $m^2_{\pi;NGB;SM}=0$;

The GMRP gives a precise interpretation of Eq. (3C.1) and $V^{\text{Re}\,normalized;1-Loop;\Lambda^2}_{SM}$ in Eq. (3B.12):

$$V^{\text{Re}\,normalized;1-Loop;\Lambda^2}_{SM} = \lambda^2\left[\frac{1}{2}h^2+\frac{1}{2}\pi_3^2+\pi_+\pi_-+\langle H\rangle h\right]^2 + m^2_{\pi;NGB;SM}\left[\frac{1}{2}h^2+\frac{1}{2}\pi_3^2+\pi_+\pi_-\right]; \quad (3C.2)$$

$$m^2_{\pi;NGB;SM}=\delta\mu^2-C^{1-Loop;\Lambda^2}_{SM}\Lambda^2;$$

**A 1-loop-induced mass-squared $m^2_{\pi;NGB;SM}$ for the three NGB (i.e. the would-be pre-Higgs-mechanism longitudinal components of massive $W^\pm_\mu, Z_\mu$) has appeared in the SM!** But it is only an artefact of not-having-yet properly enforced SSB and the Goldstone theorem. B.W.Lee and K.Symanzik's renormalization of un-gauged $O(4)L\Sigma M$ in the $\langle H\rangle$ vs. $m^2_\pi/\lambda^2$ half-plane [11,12], Appendix 2 and Ref. [10] point to the correct SM GMRP by proving that:

1. When calculations <u>begin</u> with the bare Goldstone-mode Lagrangian in Eq. (2D.7), an analogous 1-loop-induced finite NGB mass-squared $m^2_{\pi;NGB;L\Sigma M}$ appears in Goldstone-mode-$O(4)L\Sigma M$;
2. $m^2_{\pi;NGB;L\Sigma M}$ is only an artefact of using the bare Goldstone-mode Lagrangian;
3. $m^2_{\pi;NGB;L\Sigma M}$ is properly controlled by Lee/Symanzik's Higgs VSC and Goldstone SRC, which force $m_{\pi;NGB;L\Sigma M}=0$ identically, with exactly zero remnant;

Comparison of NGB masses-squared in Eqs. (3B.5,3C.2, 2E.2,2E.4) shows that

$$m^2_{\pi;NGB;SM} \xrightarrow[g_1,g_2,g_3\to 0]{} m^2_{\pi;NGB;L\Sigma M+SMQuarks\&Leptons} \quad (3C.3)$$

smoothly. $m^2_{\pi;NGB;SM}$ is therefore the SM analogy of the physical renormalized pion mass appearing in Lee/Symanzik's generic $O(4)L\Sigma M$ theories, but before properly taking the Goldstone mode limit: i.e. lingering in the $\langle H\rangle - m^2_\pi/\lambda^2$ half-plane before properly moving onto the $m^2_\pi=0$ line [10]. Therefore, in order that renormalization of Goldstone-mode-$O(4)L\Sigma M$ and the SM be consistent in the appropriate zero-gauge-coupling limit, $m^2_{\pi;NGB;SM}$ must properly be controlled by imposing Lee/Symanzik's two renormalization conditions.

Practical Implementation of GMRP for SM UV-QD:
1. <u>Goldstone SRC</u>: Set $m^2_{\pi;NGB;SM}$ to zero identically, with zero finite remnant

$$0 = m^2_{\pi;NGB;SM}\left(\frac{1}{2}h^2+\frac{1}{2}\pi_3^2+\pi_+\pi_-+\langle H\rangle h\right); \quad (3C.4)$$

$$0 = m^2_{\pi;NGB;SM} = \delta\mu^2 - \left(\frac{(1-n)(M_Z^2+2M_W^2)-3m_h^2+4\sum_{Quarks}^{Flavor,Color}m^2_{Quark}+4\sum_{Leptons}^{Flavor}m^2_{Lepton}}{16\pi^2\langle H\rangle^2}\right)\Lambda^2; \quad (3C.5)$$

$$m_h^2 = 2\lambda^2\langle H\rangle^2;$$

thus **restoring gauge invariance** and determining $\delta\mu^2$ to 1-loop in the SM

2. <u>Higgs VSC</u>: Eq. (3B.10) disallows the Higgs from simply disappearing into the vacuum.

**After GMRP, the 1-loop-UV-QD-improved scalar sector effective SM Lagrangian**, which includes all 1-loop SM UV-QD (but ignores logarithmic divergences, finite contributions and vacuum energy/bubbles) **is gauge invariant** and can be written

$$L_{SM;\Phi-Sector}^{Effective;1-Loop;\Lambda^2} = -|D_\mu \Phi|^2 - V_{SM}^{Renormalized;1-Loop;\Lambda^2};$$

$$V_{SM}^{Renormalized;1-Loop;\Lambda^2} = \lambda^2 \left[ \Phi^\dagger \Phi - \frac{1}{2}\langle H \rangle^2 \right]^2; \quad (3C.6)$$

$$H = h + \langle H \rangle; \quad \langle h \rangle = 0;$$

Eq. (3C.6) leaves the Eq. (3A.5) relation $m_h^2 / M_W^2 = 8\lambda^2 / g_2^2$ with at worst logarithmically divergent corrections, a crucial SM requirement because $\langle H \rangle^2$ is not a free parameter in the SM, but was long ago set experimentally by the HPEW muon decay constant $G_{Muon}^{Experimental}$ (see Eq. (3A.5,3A.6) and Appendix 5). 1-loop SM UV-QD are not absorbed into bare or renormalized parameters because there are no surviving finite remnants of UV-QD in $L_{SM;\Phi-Sector}^{Effective;1-Loop;\Lambda^2}$. Appendix 1 shows that Eq. (3C.6) is independent of UV regularization scheme (e.g. $n$-dimensional or Pauli-Villars cut-off).

A crucial observation of this paper is that all 1-loop UV-QD contributions to the SM Higgs' mass (i.e. the 1$^{st}$ term on the right hand side of Eq. (3C.4))

$$0 = \left( \delta\mu^2 - C_{SM}^{1-Loop;\Lambda^2} \Lambda^2 \right)\left( \frac{1}{2}h^2 \right) = m_{\pi;NGB;SM}^2 \left( \frac{1}{2}h^2 \right); \quad (3C.7)$$

also vanish identically. We are free to set $m_h^2, \langle H \rangle^2 \ll \Lambda^2$. Beyond the kinetic term and 3-point and 4-point interactions, $L_{SM;\Phi-Sector}^{Effective;1-Loop;\Lambda^2}$ gives the sensible, at worst logarithmically divergent Higgs mass

$$L_{SM;\Phi-Sector}^{Effective;1-Loop;\Lambda^2} \sim -\frac{1}{2}m_h^2 h^2 \quad m_h^2 = 2\lambda^2 \langle H \rangle^2; \quad (3C.8)$$

**Since $\langle H \rangle$ is not UV-QD [5,Appendix 5] or fine-tuned, $m_h^2$ is not fine-tuned. Fine-tuning is therefore un-necessary for weak scale $G_{Muon}^{Experimental}\langle H \rangle^2, G_{Muon}^{Experimental} m_h^2 \sim O(1)$ and the SM avoids any 1-loop HFTP.**

After Lee/Symanzik's two GMRP conditions are imposed, SM NGB mass-squared $m_{\pi;NGB;SM}^2$ are zero, SM gauge symmetry will be spontaneously broken in Eq. (3C.6), and the Higgs mechanism can do its work in the SM gauge boson and fermion sectors. We are reminded that correct renormalization of the SM is often subtle. In particular, self-consistent renormalization of the SM requires that the scalar-sector UV-QD-corrected effective Lagrangians of the SM and Goldstone-mode-$O(4)L\Sigma M$ become smoothly identical in the appropriate zero-gauge-coupling limit. It follows that, together with Higgs VSC tadpole renormalization, the UV-QD-corrected SM requires enforcement of SSB and the Goldstone theorem (essentially by hand) via imposition of Lee/Symanzik's Goldstone-SRC. Our Higgs no-fine-tuning result is therefore simply another (albeit un-familiar) consequence of the Goldstone theorem: **1-loop UV-QD contributions to the Higgs mass vanish identically and the 1-loop SM does not suffer a HFTP**.

Summary of 1-loop SM results shown by explicit calculation: Including all 1-loop SM UV-QD, but ignoring logarithmic divergences, finite parts and vacuum energy/bubbles:
1. The reader is reminded that:
    - All 1-loop SM UV-QD, in 4-fermion HPEW processes (with mass-less external fermions) mediated by gauge bosons, were shown long ago to cancel among themselves, without absorption into renormalized or bare parameters [5,Appendix 5]. **UV-QD do not arise in 1-loop HPEW**;
    - The fact that all 1-loop SM UV-QD in the Higgs' self-energy cancel after tadpole renormalization has been known [6] for more than 3 decades;
    - The bare SM VEV, $\frac{1}{2}\langle H_{Bare}\rangle^2 = \frac{1}{4\sqrt{2}G_{Muon}^{Experimental}}$ (true up to logarithmically divergent and finite corrections) is not a free parameter, but was long ago determined by the HPEW experimental muon lifetime [5,Appendix 5];
2. All 1-loop UV-QD which arise in the SM S-Matrix cancel exactly among themselves: i.e. for <u>each</u> virtual multiplet, with a subtlety in the gauge-ghost-NGB sector. There is no possibility (or need) in the SM to cancel 1-loop UV-QD between virtual bosons and fermions;
3. Explicit calculation shows that the 1-loop UV-QD in $O(4)L\Sigma M$ are the continuous smooth zero-gauge-coupling limit of those in the SM;
4. Appendix 1 shows that our 1-loop UV-QD results do not depend on choice of $n$-dimensional or Pauli-Villars cut-off regularization scheme;

Summary of 1-loop SM results in Section 3C, which we show can (almost certainly) be extended to all perturbative QCD and electroweak loop-orders in Section 3D and Appendix 6: Including all SM UV-QD, but ignoring logarithmic divergences, finite parts and vacuum energy/bubbles:

5. Neither $\langle\Phi_{Bare}\rangle^2$ nor the Higgs mass-squared $m_h^2$ can be used to absorb UV-QD arising in SM 2-point functions: they must cancel among themselves, leaving the relation $m_h^2/M_W^2 = 8\lambda^2/g_2^2$ with at worst logarithmically divergent corrections;
6. Before being fixed by HPEW, $\langle H\rangle$ is a free parameter in the SM, un-determined by the internal self-consistency of the theory (see Appendix 5). We are free to choose its numerical value as demanded by HPEW experiment $\sqrt{2}\langle H\rangle^2 G_{Muon}^{Experimantal} \approx 1$;
7. UV-QD in the SM S-Matrix cancel identically, with exactly zero finite remnant; This is traced to careful definition of the vacuum after SSB, Higgs VSC tadpole renormalization and Goldstone SRC enforcement of the Goldstone theorem;
8. Although the scale of the coefficient of the dimension-2 scalar 2-point relevant SM operators remain natural, $\delta\mu^2 \sim \Lambda^2$, no fine-tuning, even with Planck-scale UV cut-off, is necessary for $m_h^2, \langle H\rangle^2 << \Lambda^2$;
9. All relevant dimension-2 operators in the SM form a renormalized Higgs potential which is well-defined and is minimized, without fine-tuning, at the loop-corrected Higgs VEV $\langle H\rangle << \Lambda$;
10. Since, in the zero-gauge-coupling limit, UV-QD in the SM and Goldstone-mode-$O(4)L\Sigma M$ with SM quarks and leptons are the same, they must be treated the same way. Consistent renormalization requires the scalar-sector effective Lagrangians of the SM and Goldstone-mode $O(4)L\Sigma M$ to be smoothly identical (i.e. no fine-tuning discontinuity, Appendix 2) in the appropriate zero-gauge-coupling limit;

11. An artefact, in intermediate stages of SM calculations (i.e. of not having yet enforced the Goldstone theorem), is a loop-induced mass-squared $m^2_{\pi;NGB;SM}$, i.e. a finite mass for the three SM NGB. It is the SM analogy of the non-zero physical renormalized pion pole mass appearing in Lee and Symanzik's generic $O(4)L\Sigma M$ before taking the Goldstone mode limit: i.e. lingering in the $\langle H \rangle - m^2_\pi / \lambda^2$ half-plane before moving onto the $m^2_\pi = 0$ line [10];

12. All remnants of SM UV-QD are absorbed into the NGB mass-squared $m^2_{\pi;NGB;SM}$ and vanish in the SSB limit: i.e. $m^2_{\pi;NGB;SM} \to 0$;

13. The self-consistent SM GMRP carries the explicit instruction to impose Lee and Symanzik's two conditions on the SM vacuum and excited states:
    - <u>Higgs VSC</u>: the Higgs must not simply disappear into the vacuum;
    - <u>Goldstone SRC</u>: the mass-squared of SM NGB must be exactly zero $m^2_{\pi;NGB;SM} = 0$;

14. After Goldstone SRC insistence that $m^2_{\pi;NGB;SM} = 0$, the 3 mass-less NGBs are absorbed into longitudinal components of the 3 massive SM gauge bosons via the Higgs mechanism;

15. Since UV-QD are not absorbed into SM renormalized or bare parameters, no fine-tuning is necessary to enforce weak-scale $m^2_h G^{Experimantal}_{Muon}$, $M^2_W G^{Experimantal}_{Muon} \sim O(1)$, if experiment so demands;

16. The UV-QD-loop-corrected SM does not suffer a HFTP. Our no-fine-tuning-theorem for a weak-scale SM Higgs mass is simply another (albeit un-familiar) consequence of the Goldstone theorem, an exact property of the SM vacuum and excited states;

17. Since SSB SM symmetries and the Goldstone theorem are sufficient to protect a weak-scale Higgs bare or renormalized mass, and ensure that it has no HFTP, it is un-necessary to impose any new BSM symmetries;

**3D: The SM (almost certainly) has no surviving remnant of multi-loop UV-QD and does not suffer HFTP at <u>any</u> loop-order of perturbation theory**

The reader should worry that cancellation of 1-loop UV-QD is insufficient to demonstrate that the SM does not require fine-tuning. UV-QD certainly appear at multi-loop orders and fine-tuning $\delta\mu^2$ might yet be required. If each loop order contributes a factor $\hbar/16\pi^2 \sim 10^{-2}$, cancellation of UV-QD remnants to ~16 1PI loops is required to defeat a factor of $\left(G^{Experimental}_{Muon} M^2_{Planck}\right)^{-1} \sim 10^{-32}$. Ref. [10] demonstrates just such an exact cancellation for Goldstone-mode-$O(4)L\Sigma M$ and shows that 1PI multi-loop fine-tuning is un-necessary.

Appendix 6 shows that all SM S-Matrix UV-QD, including all-loop-orders perturbative QCD and electroweak corrections, (almost certainly) vanish identically. It follows closely the reasoning in Sections 2D, 3C and Ref. [10]. Appendix 6 first observes that $\langle H \rangle$ absorbs no UV-QD during the HPEW renormalization of its value to the experimental muon lifetime. Then Lee/Symanzik's two renormalization conditions are imposed:

1. <u>Higgs VSC</u>: the Higgs must not disappear into the vacuum. After gauge-dependent tadpole renormalization, a certain sub-set of operators arising in $V_{SM}^{\text{Renormalized;All-Loop};\Lambda^2}$ are still not gauge invariant;
2. <u>Goldstone SRC</u>: the mass-squared of SM NGBs must vanish identically. Gauge and BRST [21] invariance are restored to the scalar-sector SM effective Lagrangian (including all SM S-Matrix UV-QD, but ignoring logarithmic divergences and finite terms);

$$L_{SM;\Phi-\text{Sector}}^{\text{Effective;All-Loop};\Lambda^2} = -|D_\mu \Phi|^2 - V_{SM}^{\text{Renormalized;All-Loop};\Lambda^2};$$

$$V_{SM}^{\text{Renormalized;All-Loop};\Lambda^2} = \lambda^2 \left[\frac{1}{2}h^2 + \frac{1}{2}\pi_3^2 + \pi_+\pi_- + \langle H \rangle h\right]^2 = \lambda^2 \left(\Phi^\dagger \Phi - \frac{1}{2}\langle H \rangle^2\right)^2; \qquad (3D.3)$$

$$m_h^2 = 2\lambda^2 \langle H \rangle^2;$$

which gives a sensible, at worst logarithmically divergent, and *not-fine-tuned* Higgs mass.

**The reader is warned that a rigorous all-loop-order mathematical proof requires work beyond that shown in Appendix 6.**

We are free to set $G_{\text{Muon}}^{\text{Experimental}} m_h^2 \sim O(1)$ if so required by experiment. The all-loop-order relation, an extension of the 1-loop result Eq. (3A.5) to include all SM UV-QD, still reads $m_h^2 / M_W^2 = 8\lambda^2 / g_2^2$ and receives at worst logarithmically divergent and finite corrections. That fact is crucial, because the Higgs VEV-squared $\langle H \rangle^2$ is not a free parameter in the SM, but is set by the experimental values of $M_W^2, M_Z^2, G_{\text{Muon}}^{-1} \ll \Lambda^2$ [5]: no fine-tuning or absorption of UV-QD into either bare parameters, or renormalized physical input parameters, is necessary or even possible in the SM. **Therefore, the SM (almost certainly) does not suffer a UV-QD-generated HFTP at any loop-order of electro-weak and QCD perturbation theory**.

To completely avoid any fine-tuning in the SM, we must address one last detail. The reader should worry that, for an arbitrary particle spectrum (e.g. with large mass-splitting within a SM multiplet), it might still be necessary to fine-tune <u>finite</u> effects of large masses. But in the SM, with Higgs mass below the upper limit derived from comparison of HPEW 1-electroweak-loop (and bremsstrahlung) corrected lepton forward-backward and polarization asymmetries [5,18,7] with high precision LEP/SLC data [9], we have

$$G_{\text{Muon}}^{\text{Experimental}} m_j^2 \leq O(1); \quad ; \qquad (3D.4)$$
$$j = 3\text{GenerationsQuarks \& Leptons, GaugeBosons, Higgs}$$

so no such fine-tuning is necessary.

In summary, including all SM UV-QD, but ignoring logarithmic divergences, finite parts and vacuum energy/bubbles (almost certainly):
1. All UV-QD cancel identically in the SM S-matrix. This is true not only in all-loop-order electroweak perturbation theory, but also includes all-loop-order QCD perturbative corrections: the resulting $SU(3)_{\text{Color}} \times SU(2)_{\text{Isospin}} \times U(1)_{\text{Hypercharge}}$ all-loop-order SM S-Matrix has exactly zero remnant of perturbative UV-QD;
2. "Goldstone gauge" in Appendix 6 uses global BRST symmetries to manifestly confine any possible UV-QD to the SM scalar sector;

3. It is un-necessary to fine-tune remnant <u>finite</u> effects for the known SM particle spectrum, such as the top quark or Higgs masses;
4. The 1-loop SM results listed at the end of Section 3C are extended to the all-loop-orders SM;
5. Spontaneously broken SM symmetries, and the Goldstone theorem, are already sufficient to protect the bare and renormalized Higgs mass and ensure that the SM has no Higgs Fine-Tuning Problem.

**4. Conclusions: Higgs-VSC and Goldstone-SRC, rather than imposition of new BSM symmetries. Implications for the discovery potential of the LHC.**

Belief in a 1-loop HFTP in Goldstone mode $O(4)L\Sigma M$ and the SM, *previously identified as fatal mathematical flaws*, is simply mistaken. It disagrees with established renormalization theory ~ 1970, and with high precision electroweak physics (HPEW) in the LEP/SLC era: both literatures are established and proven foundations of modern mainstream theoretical physics. Ironically, that widespread belief has fruitfully formed much of the original motivation for certain proposed BSM physics. A partial list would include:
- Low energy SUSY. Commentary appears in Refs. [10,22];
- Technicolor. Commentary appears in Ref. [10];
- Pseudo-Nambu-Goldstone Higgs, Higgs-less SM;
- Little Higgs', Constraints among SM particle masses;
- Gauge-singlet scalars, Mirror fermions;
- Lee-Wick higher derivative SM;
- Conformal symmetry with Higgs of effective-dimension $(1+\varepsilon)$;
- Landscape multi-verse environmental selection of fine-tuned weak scale;

But, because it spontaneously breaks an embedded $O(4)$ symmetry, with three exactly massless NGB, the SM finds a loophole in that "fatal flaw" reasoning and avoids its own demise. Instead, the reason SM S-Matrix UV-QD sum exactly to zero is traced to correct renormalization of the SM consistent with the GMRP and SSB. In particular, we insist that self-consistent renormalization of the 1-loop SM requires that the scalar-sector UV-QD-corrected effective Lagrangians of the SM and Goldstone-mode-$O(4)L\Sigma M$ become smoothly identical (i.e. without fine-tuning discontinuity, as defined in Appendix 2) in the appropriate zero-gauge-coupling limit. It follows that, after VSC tadpole renormalization, the 1-loop UV-QD-corrected SM requires explicit enforcement of the Goldstone theorem, by imposition of Lee/Symanzik's Goldstone-SRC. 1-loop SM UV-QD, and their finite remnants, then all vanish identically. Our 1-loop no-fine-tuning-theorem for a weak-scale SM Higgs mass is therefore simply another (albeit un-familiar) consequence of the Goldstone theorem, an exact property of the SM vacuum and excited states. **It is un-necessary to impose any new BSM symmetries: the SSB 1-loop SM already has sufficient symmetry (i.e. using the Goldstone theorem) to protect the bare and renormalized Higgs masses and avoid any 1-loop HFTP**.

Although the <u>stand-alone</u> 1-loop SM, with SM loop integrals cut off at some much higher energy UV scale $\Lambda$, is insensitive to that higher scale (up to terms $\sim \ln\Lambda$), that may be spoiled if the SM is embedded/integrated into some higher scale BSM theory. But that is not the SM's problem. Given its extra-ordinary theoretical and experimental success, the Standard Model should be regarded as the resurgent new standard of naturalness and scalar no-fine-tuning excellence to which proposed BSM quantum field theories are to be compared.

Mistaken belief in huge non-vanishing remnants of cancelled UV-QD, and a HFTP, in the 1-loop SM has also historically driven an expectation that some sort of new BSM physics <u>must</u> appear at low-energy ($\leq 14$ TeV) and at the LHC. But the Standard Model (i.e. with modified neutrino sector and coupled to <u>classical</u> general relativity) is the most powerful, accurate, predictive, successful and experimentally verified scientific theory known to humans. Although it is sometimes necessary to dig more deeply, we should not be surprised by the SM's extra-ordinary explanatory power at the quantum field theoretic level. This includes its spectacular successful prediction of the top quark mass by the HPEW comparison of 1-electroweak-loop corrected total cross sections, lepton forward-backward and inclusive electron-polarization asymmetries on/near $Z_\mu$ resonance [5,8] with high-precision LEP1/SLC experiments [9]. We have shown here that the SM finds a SSB loophole through which all 1-loop SM UV-QD remnants vanish identically. Widespread reports of the SM's presumed HFTP death are incorrect and our results re-open the very real possibility that the discovery potential of the LHC might be confined to SM physics. Although it currently lacks a credible dark matter candidate [23], the SM solves the phenomenological problems of many BSM models: e.g. "*automatic conservation of baryon and lepton number in interactions up to dimension 5 and 4, respectively; natural conservation of flavors in neutral currents; a small neutron electric dipole moment* [24]" and constraints from HPEW physics. The crucial make/break test (i.e. a win/win experiment!) for the SM is LHC discovery/exclusion of the SM Higgs with mass below the predicted upper limit from HPEW comparison of 1-electroweak-loop-corrected total cross sections, forward-backward asymmetries to leptons and inclusive electron-polarization asymmetries on/near $Z_\mu$ resonance [5,7,18] against high precision LEP1/SLC experiments [9].

## <u>Acknowledgements</u>

I thank with great pleasure the Mitchell Institute for Fundamental Physics & Astronomy at Texas A&M University, where this work was done, and its Director Christopher Pope. I thank Jonathan Butterworth and the Department of Physics at University College London for support. I thank Luis Alvarez-Guame and the CERN Theoretical Physics Division for hospitality in 2009-2010, where this work began. I thank Bruce Campbell, Savas Dimopoulos, Jonathan R. Ellis, Katherine Freese, Amir E. Mossafa, Michael Peskin, Dmitry Podolsky, Chris Pope, Ergin Sezgin, Glenn Starkman and Jennifer Thomas for valuable discussions. I am indebted to Robert B. Laughlin for his private communication that stability conditions in solid state physics, a physical property characterizing phases of matter, are known, in certain cases, to cause UV divergences to sum automatically to zero. I am indebted to Raymond Stora for reconstructing the history of the renormalization of un-gauged $O(4)L\Sigma M$, and for insisting on K.Symanzik's "new testament" dictum (at the beginning in Section 2).

## <u>References</u>

operators with 1-loop UV-QD coefficients arising in the SM would read (i.e. if taken at face value and written in our language):

$$L_{SM;\Phi-Sector}^{1-Loop;\Lambda^2}\Big|_{\substack{MVeltman \\ ActaPhysPol \\ B12(1981)437}} = \left(-\frac{1}{2}h^2 + \frac{1}{2}\pi_3^2 + \pi_+\pi_- - \langle H \rangle h\right)$$

$$* \Lambda^2 \left(\frac{(1-n)(M_Z^2 + 2M_W^2) - 3m_h^2 + 4\sum_{Quarks}^{Flavor,Color} m_{Quark}^2 + 4\sum_{Leptons}^{Flavor} m_{Lepton}^2}{16\pi^2 \langle H \rangle^2}\right); \quad (R.1)$$

Comparison with our quadratically divergent SM symmetric counter-term Eq. (3B.6)

$$L_{SM;\Phi-Sector}^{CounterTerm;\Lambda^2}\Big|_{Symmetric} = -\delta\mu^2\left(+\frac{1}{2}h^2 + \frac{1}{2}\pi_3^2 + \pi_+\pi_- + \langle H \rangle h\right) \quad (R.2)$$

reveals that Eq. (R.1) is not $O(4)$ invariant before SSB: i.e. because of its relative minus sign (with which we disagree) between the Higgs self-energy 2-point operator and that of the Nambu-Goldstone Bosons. In contrast, our analogous expression Eq. (3B.4) and counter-term Eq. (3B.6) are manifestly $O(4)$ invariant before SSB, as they must be in the SM. If the relative sign in Eq. (R.1) is not a typographical error, it is impossible to cancel exactly all 1-loop SM UV-QD between Eqs. (R.1) and (R.2);

**Appendix 1: Representation of 1-loop UV-QD in *n*-dimensional regularization mapped onto Pauli-Villars regularization**

M.J.G.Veltman proved more than 30 years ago [2] that the Passarino & Veltman $A(m^2)$ and $B_{22}(q^2, m_1^2, m_2^2)$ functions [3] can be made to properly capture and represent UV-QD in the SM (if certain subtleties are obeyed) in both $n$-dimensional regularization (i.e. with poles at $n=2$) and Pauli-Villars with an ultra-violet cut-off $\Lambda$:

$$A(m^2) = \int \frac{d^n k}{i\pi^2} \frac{1}{(k^2 + m^2)};$$

$$B_{\mu\nu}; B_\mu; B_0 = \int \frac{d^n k}{i\pi^2} \frac{k_\mu k_\nu; k_\nu; 1}{(k^2 + m_1^2)((k+q)^2 + m_2^2)} \quad (A1.1)$$

$$B_{\mu\nu} = \delta_{\mu\nu} B_{22} + q_\mu q_\nu B_{21}; \quad B_\mu = q_\mu B_1; \quad B_3 = B_{21} + B_1;$$

On the other hand, these 1-loop integrals can be written with a UV cut-off $\Lambda$ [2]:

$$\frac{2}{4-n} - \gamma - \ln\pi \xrightarrow{n \to 4} \ln\Lambda^2;$$

$$nB_{22}(q^2, m^2, m^2) - A(m^2) \xrightarrow{n \to 4} \left(-\frac{1}{3} q^2 - m^2\right) \ln\left(\frac{\Lambda^2}{m^2}\right) + \textit{finite};$$

$$A(m^2) \xrightarrow{n \to 4} \Lambda^2 - m^2 \left(\ln\frac{\Lambda^2}{m^2} + 1\right);$$

$$(2-n)A(m^2) \xrightarrow{n \to 4} \sim m^2 \ln\left(\frac{\Lambda^2}{m^2}\right) + \textit{finite};$$

(A1.2)

$$B_0(q^2,m_1^2,m_2^2) \xrightarrow{n\to 4} \ln\Lambda^2 - \int_0^1 dx \ln\left[m_1^2 + \left(q^2 + m_2^2 - m_1^2\right)x - x^2 q^2 - i\varepsilon\right];$$

$$B_1(q^2,m_1^2,m_2^2) \xrightarrow{n\to 4} -\frac{1}{2}\ln\Lambda^2 + \int_0^1 dx\, x \ln\left[m_1^2 + \left(q^2 + m_2^2 - m_1^2\right)x - x^2 q^2 - i\varepsilon\right]; \qquad (A1.3)$$

$$B_{21}(q^2,m_1^2,m_2^2) \xrightarrow{n\to 4} \frac{1}{3}\ln\Lambda^2 - \int_0^1 dx\, x^2 \ln\left[m_1^2 + \left(q^2 + m_2^2 - m_1^2\right)x - x^2 q^2 - i\varepsilon\right];$$

The weak demand is then made, i.e. of any well-defined UV regularization scheme, that one can self-consistently change variables of integration. 1-loop UV-QD then cancel exactly (with zero finite remnant) in expressions where they do not truly arise: e.g.

$$\delta_{\mu\nu}B_{\mu\nu} + m_1^2 B_0 - A(m_2^2) = \int \frac{d^n k}{i\pi^2}\left(\frac{\delta_{\mu\nu}k_\mu k_\nu + m_1^2}{\left(k^2 + m_1^2\right)\left((k+q)^2 + m_2^2\right)} - \frac{1}{\left(k^2 + m_2^2\right)}\right) = 0; \qquad (A1.4)$$

The finite parts of $A(m^2)$ in Eq. (A1.2), and $B_{22}(q^2,m_1^2,m_2^2)$, are fixed so that changing integration variables is consistent with the expressions for $B_0(q^2,m_1,m_2)$, etc. in Eq. (A1.3), and more generally that expressions where UV-QD do not arise, such as

$$A(m_1^2) - A(m_2^2) = \left(m_2^2 - m_1^2\right)B_0(q^2,m_1^2,m_2^2); \qquad (A1.5)$$

are true. M.J.G.Veltman also argued that a SUSY extension of $n$-dimensional regularization [25] allows relaxation of the last of Eqs. (A1.2) for the virtual $Z_\mu, W_\mu^\pm$ contributions $p_{8A}, p_{8B}$ in Eqs. (3B.5, A3.15, R.1) and setting $n = 4$ there. Whether $n = 4$ (or $n = 2$ in the SM as Eq. (A1.2) insists) in Eq. (3B.1) does not matter for this paper. The vanishing of UV-QD in Goldstone-mode un-gauged $O(4)L\Sigma M$ and the SM does not therefore depend on whether $n$-dimensional regularization or Pauli-Villars UV cut-off regularization is used

## Appendix 2: The Goldstone Mode UV-QD Renormalization Prescription (GMRP) and "Fine-tuning discontinuity" in un-gauged $O(4)L\Sigma M$

In order to better understand the correct self-consistent renormalization prescription for the SM, we re-do the calculations and analysis of Section 2D, but <u>beginning</u> with the Goldstone-mode-$O(4)L\Sigma M$ bare Lagrangian, i.e. Eq. (2A.1) with $\varepsilon = 0$.

$$L_{L\Sigma M}^{Bare;\Lambda^2} = -\left|\partial_\mu \Phi\right|^2 - \lambda^2\left(\Phi^\dagger\Phi - \frac{1}{2}\langle H\rangle^2\right)^2 + L_{L\Sigma M}^{CounterTerm;\Lambda^2};$$

$$L_{L\Sigma M}^{CounterTerm;\Lambda^2} = -\delta\mu^2\left(\Phi^\dagger\Phi - \frac{1}{2}\langle H\rangle^2\right); \qquad (A2.1)$$

where we ignore vacuum energy/bubbles. Using this bare Lagrangian and the 1-loop UV-QD result in Eq. (2B.1, 2B.2), we form the 1-loop-UV-QD-improved effective Goldstone mode Lagrangian, which includes all scalar 2-point 1-loop UV-QD self-energies and 1-point 1-loop UV-QD tadpoles, but ignores 1-loop logarithmically divergent, finite contributions and vacuum energy/bubbles:

$$L_{GoldstoneMode;L\Sigma M}^{Effective;1-Loop;\Lambda^2} = L_{GoldstoneMode;L\Sigma M}^{Bare} + L_{L\Sigma M}^{1-Loop;\Lambda^2}$$
$$= -\left|\partial_\mu \Phi\right|^2 - V_{GoldstoneMode;L\Sigma M}^{Renormalized;1-Loop;\Lambda^2}; \qquad (A2.2)$$

$$V_{GoldstoneMode;L\Sigma M}^{\text{Re}normalized;1-Loop;\Lambda^2} = \lambda^2 \left[\frac{1}{2}h^2 + \frac{1}{2}\pi_3^2 + \pi_+\pi_- + \langle H \rangle h\right]^2 \qquad (A2.3)$$
$$+ m_{\pi;NGB;L\Sigma M}^2 \left[\frac{1}{2}h^2 + \frac{1}{2}\pi_3^2 + \pi_+\pi_- + \langle H \rangle h\right];$$

A 1-loop-induced finite (after cancellation of UV-QD) mass-squared $m_{\pi;NGB;L\Sigma M}^2$ for the three Nambu-Goldstone Bosons, has appeared!

$$m_{\pi;NGB;L\Sigma M}^2 = \delta\mu^2 + \left(\frac{3m_h^2}{16\pi^2 \langle H \rangle^2}\right)\Lambda^2; \qquad (A2.4)$$

but, as shown in Section 2D, it is only an artefact of not having yet properly enforced the Goldstone theorem: i.e. a contribution to the physical renormalized pion mass appearing in Lee/Symanzik's generic $O(4)L\Sigma M$ theories, before taking the Goldstone mode limit; i.e. in the $\langle H \rangle - m_\pi^2/\lambda^2$ plane before moving onto the $m_\pi^2 = 0$ line [11,12].

Section 2D and Ref. [10] prove that the correct self-consistent **GMRP** for UV-QD in Goldstone-mode-$O(4)L\Sigma M$ is to impose Lee/Symanzik's two conditions

1. Goldstone SRC: $\delta\mu^2$ is used to set $m_{\pi;NGB;L\Sigma M}^2$ to zero identically, with exactly zero finite remnant, and restore axial current conservation;
2. Higgs VSC: After Goldstone SRC, it is no longer necessary to cancel tadpoles

We then set $\langle H \rangle^{NoFineTuning}$ to the experimental value, its name emphasizing that fine-tuning is un-necessary [11,12, Appendix 4] for a weak scale $G_{Muon}^{Experimwental} \left(\langle H \rangle^{NoFineTuning}\right)^2 \sim O(1)$ and recover Eq. (2D.2):

$$V_{GoldstoneModeL\Sigma M}^{\text{Re}normalized;1-Loop;\Lambda^2} = \lambda^2 \left[\Phi^\dagger\Phi - \frac{1}{2}\left(\langle H \rangle^{NoFineTuning}\right)^2\right]^2; \qquad (A2.5)$$
$$m_{\pi;NGB;L\Sigma M}^2 = 0; \quad H = h + \langle H \rangle^{NoFineTuning}; \quad \langle h \rangle = 0;$$

The GMRP clearly generalizes, to include SM quarks and leptons, and all-loop-orders, in Goldstone-mode-$O(4)L\Sigma M$ [10].

Because the *tree-level* Standard Model is already in "Goldstone mode", the same subtlety arises in the 1-loop SM. After cancellation of UV-QD, a finite mass-squared $m_{\pi;NGB;SM}^2$ for the three NGBs appears in Eq. (3C.2) as a 1-loop-induced artefact in the SM! Since, in the zero-gauge-coupling limit, the SM UV-QD analysis must be identical to that of Goldstone-mode-$O(4)L\Sigma M$, Sections 2D, 3C, this Appendix and Ref. [10] insist that self-consistent renormalization of the SM requires imposition of Lee/Symanzik's Goldstone SRC and Higgs VSC in the SM. The UV-QD GMRP enforces SSB, the Goldstone theorem, and $m_{\pi;NGB;SM}^2 = 0$ identically, with zero surviving remnant, in the 1-loop-UV-QD-improved SM S-Matrix. Section 3D and Appendix 6 extends it to all electro-weak and QCD loop-orders.

**Fine-tuning discontinuity:** Ignoring vacuum energy, re-write the Goldstone-mode-$O(4)L\Sigma M$ 1-loop Eq. (A2.3)

$$V_{GoldstoneMode;L\Sigma M}^{\text{Renormalized};1-Loop;\Lambda^2} = \lambda^2 \left[ \frac{1}{2}(h+\langle H \rangle)^2 + \frac{1}{2}\pi_3^2 + \pi_+\pi_- - \frac{1}{2}\left(\langle H \rangle^2 - \frac{m_{\pi;NGB;L\Sigma M}^2}{\lambda^2}\right) \right]^2$$

$$= \lambda^2 \left[ \frac{1}{2}H^2 + \frac{1}{2}\pi_3^2 + \pi_+\pi_- - \frac{1}{2}\left(\langle H \rangle^2 - \frac{m_{\pi;NGB;L\Sigma M}^2}{\lambda^2}\right) \right]^2 \quad (A2.6)$$

$$= \lambda^2 \left[ \Phi^\dagger \Phi - \frac{1}{2}\left(\langle H \rangle^{FineTuned}\right)^2 \right]^2;$$

$$\langle H \rangle^{FineTuned} = \left(\langle H \rangle^2 - \frac{m_{\pi;NGB;L\Sigma M}^2}{\lambda^2}\right)^{1/2} = \left(\langle H \rangle^2 - \frac{1}{\lambda^2}\left(\delta\mu^2 + 6\lambda^2 \frac{\Lambda^2}{16\pi^2}\right)\right)^{1/2};$$

and re-scale the Higgs VEV

$$H = h + \langle H \rangle^{FineTuned}; \quad \langle h \rangle = 0; \quad (A2.7)$$

The temptation is to regard the Higgs VEV in Eq. (A2.7) as fine-tuned. If it were, the two different approaches to renormalization, Eqs. (A2.5, A2.6), would generate a "fine-tuning discontinuity" in the 1-loop UV-QD renormalization of Goldstone-mode-$O(4)L\Sigma M$

$$\Delta_{Discontinuity}^{FineTuning} \langle H \rangle^2 = \left(\langle H \rangle^{FineTuned}\right)^2 - \left(\langle H \rangle^{NoFineTuning}\right)^2 = -\delta\mu^2 - 6\lambda^2 \frac{\Lambda^2}{16\pi^2} = -\frac{m_{\pi;NGB;L\Sigma M}^2}{\lambda^2}; \quad (A2.8)$$

even after cancellation of UV-QD in Eq. (A2.8). But Lee/Symanzik provide the self-consistent way out of this nasty discontinuity by insisting that the scalar field and Higgs VEV

$$\Phi^{Bare} = \frac{1}{\sqrt{2}}\begin{bmatrix} H + i\pi_3 \\ -\pi_2 + i\pi_1 \end{bmatrix}^{Bare}; \quad \Phi^{\text{Renormalized}} = \frac{1}{\sqrt{2}}\begin{bmatrix} H + i\pi_3 \\ -\pi_2 + i\pi_1 \end{bmatrix}^{\text{Renormalized}}; \quad (A2.9)$$

$$\Phi^{Bare} = Z_\Phi^{1/2} \Phi^{\text{Renormalized}}; \quad \langle H \rangle^{Bare} = Z_\Phi^{1/2} \langle H \rangle^{\text{Renormalized}};$$

are multiplicatively renormalized [11,12,Appendix 4] and that Goldstone mode is simply the $m_\pi^2 = 0$ line in the $\langle H \rangle - m_\pi^2/\lambda^2$ half-plane [11,12]. The GMRP, $m_\pi^2 \to m_{\pi;NGB;L\Sigma M}^2 = 0$, forces

$$\Delta_{Discontinuity}^{FineTuning} \langle H \rangle^2 = 0; \quad (A2.10)$$

to vanish identically, with exactly zero finite UV-QD remnant.

## Appendix 3: 1-loop UV-QD arise only in SM scalar sector Lagrangian $L_{SM}^{1-Loop;\Lambda^2}$

**1) Gauge invariant 2-point graphs without virtual gauge bosons or ghosts:** SM UV-QD self-energies are calculated in a general $R_\xi$ gauge. The analogous un-gauged $O(4)L\Sigma M$ diagrams are drawn in Figs. 3, 4, 5 and 7 of Ref. [10]. Using the Feynman diagram naming convention of Ref. [2] we have: $P_{9A}$ with virtual $\pi_3$; $P_{9B}$ with virtual $\pi_\pm$; $P_{10}$ with virtual $h$; $P_{11}$ with $K^{th}$ virtual SM quark; $P_{11}$ with $K^{th}$ virtual SM lepton;

$$16\pi^2 p_{9A} = -\lambda^2; \quad 16\pi^2 p_{9B} = -2\lambda^2; \quad 16\pi^2 p_{10} = -3\lambda^2; \quad (A3.1)$$

$$p_{11} = \sum_K^{SMQuarks} p_{11;K}^{Quark} + \sum_K^{SMLeptons} p_{11;K}^{Lepton}; \quad 16\pi^2 p_{11;K}^{Quark} = N_{Colors} \frac{m_K^2}{\langle H \rangle^2}; \quad 16\pi^2 p_{11;K}^{Lepton} = \frac{m_K^2}{\langle H \rangle^2}; \quad (A3.2)$$

$$L_{SM;hh}^{1-Loop;2-\text{point};\Lambda^2}\bigg|_{\substack{NoVirtual\\GaugesOrGhosts}} = \Lambda^2(p_{9A} + p_{9B} + p_{10} + p_{11})\left(\frac{1}{2}h^2\right); \tag{A3.3}$$

But the UV-QD SM contribution of <u>each</u> of the 1-loop 2-point $\pi_3\pi_3$ neutral-Nambu-Goldstone self-energy diagrams in $L_{SM;\pi_3\pi_3}^{1-Loop;2-\text{point};\Lambda^2}\bigg|_{\substack{NoVirtual\\GaugesOrGhosts}}$ and the 1-loop 2-point $\pi_-\pi_+$ charged-Nambu-Goldstone self-energy diagrams in $L_{SM;\pi_+\pi_-}^{1-Loop;2-\text{point};\Lambda^2}\bigg|_{\substack{NoVirtual\\GaugesOrGhosts}}$ is related (by explicit calculation and $O(4)$ symmetry) to its associated 1-loop 2-point $hh$ Higgs' self-energy diagram in $L_{SM;hh}^{1-Loop;2-\text{point};\Lambda^2}\bigg|_{\substack{NoVirtual\\GaugesOrGhosts}}$ by Clebsch-Gordon coefficients and combinatorics. Adding all these into the 1-loop 2-point SM scalar-sector UV-QD Lagrangian, due to graphs without virtual gauges or ghosts, gives:

$$L_{SM}^{1-Loop;2-\text{point};\Lambda^2}\bigg|_{\substack{NoVirtual\\GaugesOrGhosts}} = L_{SM;hh}^{1-Loop;2-\text{point};\Lambda^2}\bigg|_{\substack{NoVirtual\\GaugesOrGhosts}} + L_{SM;\pi_3\pi_3}^{1-Loop;2-\text{point};\Lambda^2}\bigg|_{\substack{NoVirtual\\GaugesOrGhosts}}$$
$$+ L_{SM;\pi_+\pi_-}^{1-Loop;2-\text{point};\Lambda^2}\bigg|_{\substack{NoVirtual\\GaugesOrGhosts}} = \Lambda^2(p_{9A} + p_{9B} + p_{10} + p_{11})\left(\frac{1}{2}h^2 + \frac{1}{2}\pi_3^2 + \pi_+\pi_-\right); \tag{A3.4}$$

**2) <u>Gauge invariant 1-point tadpole graphs without virtual gauge bosons or ghosts:</u>** SM UV-QD tadpoles are here re-calculated in a general $R_\xi$ gauge. The analogous un-gauged $O(4)L\Sigma M$ diagrams are drawn in Figs. 6 and 8 of Ref. [10]. Using the Feynman diagram naming convention of Ref. [2] we have: $T_{1A}$ with virtual $\pi_3$; $T_{1B}$ with virtual $\pi_\pm$; $T_4$ with virtual $h$; $T_6$ with $K^{th}$ virtual SM quark; $T_6$ with $K^{th}$ virtual SM lepton:

$$L_{SM}^{1-Loop;1-\text{point};\Lambda^2}\bigg|_{\substack{NoVirtual\\GaugesOrGhosts}} = (t_{1A} + t_{1B} + t_4 + t_6)\Lambda^2 h; \tag{A3.5}$$

But explicit calculation (and broken $O(4)$ symmetry) shows that the UV-QD in <u>each</u> of the 1-loop 1-point tadpole diagrams in $L_{SM;h\langle H\rangle}^{1-Loop;1-\text{point};\Lambda^2}\bigg|_{\substack{NoVirtual\\GaugesOrGhosts}}$ is proportional to the UV-QD in its associated 1-loop 2-point Higgs self-energy diagram in $L_{SM;hh}^{1-Loop;2-\text{point};\Lambda^2}\bigg|_{\substack{NoVirtual\\GaugesOrGhosts}}$:

$$t_{1A} = p_{9A}\langle H\rangle; \quad t_{1B} = p_{9B}\langle H\rangle; \quad t_4 = p_{10}\langle H\rangle; \tag{A3.6}$$

$$t_6 = \sum_K^{SMQuarks} t_{6;K}^{Quark} + \sum_K^{SMLeptons} t_{6;K}^{Lepton} = p_{11}\langle H\rangle; \quad t_{6;K}^{Quark} = p_{11;K}^{Quark}\langle H\rangle; \quad t_{6;K}^{Quark} = p_{11;K}^{Quark}\langle H\rangle; \tag{A3.7}$$

so that

$$L_{SM}^{1-Loop;1-\text{point};\Lambda^2}\bigg|_{\substack{NoVirtual\\GaugesOrGhosts}} = \Lambda^2(p_{9A} + p_{9B} + p_{10} + p_{11})(\langle H\rangle h); \tag{A3.8}$$

Gathering these results, we form the gauge invariant 1-loop UV-QD SM Lagrangian, which arises from SM scalar self-energy and tadpole graphs with no virtual gauge bosons or ghosts

$$L_{SM}^{1-Loop;\Lambda^2}\bigg|_{\substack{NoVirtual\\GaugesOrGhosts}} = L_{SM}^{1-Loop;2-\text{point};\Lambda^2}\bigg|_{\substack{NoVirtual\\GaugesOrGhosts}} + L_{SM}^{1-Loop;1-\text{point};\Lambda^2}\bigg|_{\substack{NoVirtual\\GaugesOrGhosts}}$$

(A3.9)

$$= C_{SM}^{1-Loop;\Lambda^2}\bigg|_{\substack{NoVirtual\\GaugesOrGhosts}} \Lambda^2\left(\frac{1}{2}h^2 + \frac{1}{2}\pi_3^2 + \pi_+\pi_- + \langle H\rangle h\right);$$

$$C_{SM}^{1-Loop;\Lambda^2}\bigg|_{\substack{NoVirtual\\GaugesOrGhosts}} = \left(\frac{-6\lambda^2\langle H\rangle^2 + 4\sum_{Quarks}^{Flavor,Color} m_{Quark}^2 + 4\sum_{Leptons}^{Flavor} m_{Lepton}^2}{16\pi^2\langle H\rangle^2}\right);$$

(A3.10)

For pedagogical clarity, these gauge invariant 1-loop SM UV-QD are usefully classified according to whether they arise from
- Virtual SM NGB and Higgs corresponding to those 1-loop UV-QD in un-gauged $O(4)L\Sigma M$ discussed in Section 2B, 2C and 2D;
- Virtual SM quarks and leptons corresponding to those 1-loop UV-QD in un-gauged $O(4)L\Sigma M$ discussed in Section 2E;

$$C_{SM}^{1-Loop;\Lambda^2}\bigg|_{\substack{NoVirtual\\GaugesOrGhosts}} = C_{SM}^{1-Loop;\Lambda^2}\bigg|_{\substack{Virtual\\NGB\&Higgs}} + C_{SM}^{1-Loop;\Lambda^2}\bigg|_{\substack{VirtualSM\\Quarks\&Leptons}}$$

$$C_{SM}^{1-Loop;\Lambda^2}\bigg|_{\substack{Virtual\\NGB\&Higgs}} \Lambda^2 = \left(-6\lambda^2\right)\frac{\Lambda^2}{16\pi^2};$$

(A3.11)

$$L_{SM}^{1-Loop;\Lambda^2}\bigg|_{\substack{Virtual\\NGB\&Higgs}} = C_{SM}^{1-Loop;\Lambda^2}\bigg|_{\substack{Virtual\\NGB\&Higgs}} \Lambda^2\left(\frac{1}{2}h^2 + \frac{1}{2}\pi_3^2 + \pi_+\pi_- + \langle H\rangle h\right);$$

(A3.12)

$$C_{SM}^{1-Loop;\Lambda^2}\bigg|_{\substack{VirtualSM\\Quarks\&Leptons}} \Lambda^2 = \left(4\sum_{Quarks}^{Flavor,Color} m_{Quark}^2 + 4\sum_{Leptons}^{Flavor} m_{Lepton}^2\right)\frac{\Lambda^2}{16\pi^2};$$

(A3.13)

$$L_{SM}^{1-Loop;\Lambda^2}\bigg|_{\substack{VirtualSM\\Quarks\&Leptons}} = C_{SM}^{1-Loop;\Lambda^2}\bigg|_{\substack{VirtualSM\\Quarks\&Leptons}} \Lambda^2\left(\frac{1}{2}h^2 + \frac{1}{2}\pi_3^2 + \pi_+\pi_- + \langle H\rangle h\right);$$

(A3.14)

**3) Gauge invariant 2-point graphs containing virtual gauge bosons or ghosts:** SM UV-QD self-energies are here re-calculated in a general $R_\xi$ gauge. Using the Feynman diagram naming convention of Ref. [2] we have: $P_{2A}$ with virtual $(\pi_3, Z_\mu)$; $P_{2B}$ with virtual $(\pi_\pm, W_\mu^\mp)$; $P_{8A}$ with virtual $Z_\mu$; $P_{8B}$ with virtual $W_\mu^\pm$:

$$16\pi^2 p_{2A} = \frac{g_2^2}{4c_\theta^2}\left(1-(\xi-1)\right); \quad 16\pi^2 p_{8A} = -\frac{g_2^2}{4c_\theta^2}\left(n-(\xi-1)\right);$$

(A3.15)

$$16\pi^2 p_{2B} = \frac{1}{2}g_2^2\left(1-(\xi-1)\right); \quad 16\pi^2 p_{8B} = -\frac{1}{2}g_2^2\left(n-(\xi-1)\right);$$

$$L_{SM;hh}^{1-Loop;2-\text{point};\Lambda^2}\bigg|_{\substack{Virtual\\GaugesOrGhosts}} = \Lambda^2\left(\{p_{2A}+p_{8A}\}+\{p_{2B}+p_{8B}\}\right)\left(\frac{1}{2}h^2\right);$$

(A3.16)

The expressions in the curly brackets are gauge invariant. The total gauge invariant 1-loop UV-QD contribution to the SM 1-loop 2-point Higgs' self-energy is:

$$L_{SM;hh}^{1-Loop;2-\text{point};\Lambda^2} = \Lambda^2 \left(p_{2A} + p_{2B} + p_{8A} + p_{8B} + p_{9A} + p_{9B} + p_{10} + p_{11}\right)\left(\frac{1}{2}h^2\right); \quad (A3.17)$$

But the UV-QD SM contribution of <u>each</u> of the 1-loop 2-point $\pi_3\pi_3$ neutral-Nambu-Goldstone self-energy diagrams in $L_{SM;\pi_3\pi_3}^{1-Loop;2-\text{point};\Lambda^2}$ and the 1-loop 2-point $\pi_-\pi_+$ charged-Nambu-Goldstone self-energy diagrams in $L_{SM;\pi_-\pi_+}^{1-Loop;2-\text{point};\Lambda^2}$ is related (by explicit calculation and $O(4)$ symmetry) to its associated 1-loop 2-point $hh$ Higgs' self-energy diagram in $L_{SM;hh}^{1-Loop;2-\text{point};\Lambda^2}$ by Clebsch-Gordon coefficients and combinatorics. Adding all these into the gauge invariant UV-QD effective 1-loop 2-point SM scalar-sector Lagrangian gives:

$$L_{SM}^{1-Loop;2-\text{point};\Lambda^2} = L_{SM;hh}^{1-Loop;2-\text{point};\Lambda^2} + L_{SM;\pi_3\pi_3}^{1-Loop;2-\text{point};\Lambda^2} + L_{SM;\pi_-\pi_+}^{1-Loop;2-\text{point};\Lambda^2}$$
$$= C_{SM}^{1-Loop;\Lambda^2} \Lambda^2 \left(\frac{1}{2}h^2 + \frac{1}{2}\pi_3^2 + \pi_+\pi_-\right); \quad (A3.18)$$

$$C_{SM}^{1-Loop;\Lambda^2} = \left(\frac{(1-n)(M_Z^2 + 2M_W^2) - 3m_h^2 + 4\sum_{Quarks}^{Flavor,Color} m_{Quark}^2 + 4\sum_{Leptons}^{Flavor} m_{Lepton}^2}{16\pi^2 \langle H \rangle^2}\right); \quad (A3.19)$$

with gauge invariant and famous [2] $C_{SM}^{1-Loop;\Lambda^2}$.

### 4) Gauge-dependent 1-point tadpole graphs containing virtual gauge bosons or ghosts:

After SSB, $\langle H \rangle^2 \geq 0$, the SM also receives UV-QD contributions from 1-loop 1-point $\langle H \rangle h$ tadpole diagrams. Although calculated, agreed [2,6] and listed in Ref. [2] in $R_\xi : \xi = 1$ gauge, we re-calculate them here in a general $R_\xi$ gauge:

- $T_{2A}$ with virtual neutral ghosts $\bar{\zeta}^Z, \eta^Z$; $T_{2B}$ with virtual charged ghosts $\bar{\zeta}^\pm, \eta^\pm$;

$$16\pi^2 t_{2A} = \frac{g_2^2}{4c_\theta^2}\langle H \rangle; \quad 16\pi^2 t_{2B} = \frac{1}{2}g_2^2\langle H \rangle; \quad (A3.20)$$

- $T_{3A}$ with virtual neutral transverse gauge boson $Z_\mu$; $T_{3B}$ with virtual charged transverse gauge bosons $W_\mu^\pm$;

$$16\pi^2 t_{3A} = -\frac{g_2^2}{4c_\theta^2}\left(n - (\xi - 1)\right)\langle H \rangle; \quad 16\pi^2 t_{3B} = -\frac{1}{2}g_2^2\left(n - (\xi - 1)\right)\langle H \rangle; \quad (A3.21)$$

The UV-QD tadpole contribution to S-matrix elements (e.g. boson and fermion propagator insertions) can be executed via a 1-loop tadpole SM Lagrangian

$$L_{SM;\langle H \rangle h}^{1-Loop;1-\text{point};\Lambda^2} = \left(t_{2A} + t_{2B} + t_{3A} + t_{3B} + t_{1A} + t_{1B} + t_4 + t_6\right)\Lambda^2 h; \quad (A3.22)$$

Since all of the diagrams, except $T_{3A}, T_{3B}$, form a gauge invariant set, it is useful to write the 1-loop SM UV-QD tadpole contributions separated into two coefficients: **gauge invariant** $C_{SM}^{1-Loop;\Lambda^2}$ and **gauge-dependent** $C_{SM;\xi}^{1-Loop;\Lambda^2}$

$$L_{SM;\langle H \rangle h}^{1-Loop;1-\text{point};\Lambda^2} = C_{SM;\xi}^{1-Loop;\Lambda^2} C_{SM}^{1-Loop;\Lambda^2} \Lambda^2 \langle H \rangle h; \quad (A3.23)$$

so the total UV-QD 1-loop SM Lagrangian is

$$L_{SM}^{1-Loop;\Lambda^2} = L_{SM;\Phi^\dagger\Phi}^{1-Loop;2-po\text{int};\Lambda^2} + L_{SM;\langle H \rangle h}^{1-Loop;1-po\text{int};\Lambda^2}$$
$$= C_{SM}^{1-Loop;\Lambda^2} \Lambda^2 \left( \frac{1}{2}h^2 + \frac{1}{2}\pi_3^2 + \pi_+\pi_- + C_{SM;\xi}^{1-Loop;\Lambda^2} \langle H \rangle h \right); \quad (A3.24)$$

Note that, in $R_\xi : \xi = 1$ gauge, $C_{SM;\xi=1}^{1-Loop;\Lambda^2} = 1$. The reader is reminded that this Appendix ignores logarithmic divergences and finite contributions. In a general $R_\xi$ gauge, construction of 1-loop gauge invariant physical results, including log divergences and finite parts, can require some care [2,3,4,5,6,7,8,9,19,20,Appendix 5] while treating:
- Vector self-energy, vertex and box contributions [5,6,16];
- Scalar self-energy and tadpole graphs with no virtual transverse gauge bosons or ghosts;
- Scalar self-energy graphs with virtual transverse gauge bosons and ghosts;
- Tadpole graphs with virtual transverse gauge bosons.
- Tadpole graphs with virtual ghosts;

**Even the UV-QD part of the sum of SM tadpole graphs is not by itself gauge invariant!**

**Appendix 4: Outline of B.W. Lee's 1970 proof that $\langle H \rangle^{Bare}$ is not UV-QD in $O(4)L\Sigma M$, receiving only logarithmic divergences from $\Phi$ wave function renormalization [11].**

- The **symmetric theory**, i.e. "Wigner mode" with $\varepsilon_{L\Sigma M} = \langle H \rangle m_\pi^2 \to 0$ with $\langle H \rangle \to 0$ holding $m_\pi^2 \neq 0$, is shown finite with $\mu_{Bare}^2$ (analogous with Eq. (2A.2) and equal for purposes of UV-QD), together with dimensionless-coupling-constant and wave function renormalization

$$(H, \vec{\pi})^{Bare} = Z_\Phi^{1/2} (H, \vec{\pi})^{\text{Re}normalized};$$
$$\lambda_{\text{Re}normalized}^2 = Z_\Phi^{1/2} Z_{\lambda^2} \lambda_{Bare}^2; \quad (A4.1)$$

- Excluding contributions from $L_{L\Sigma M}^{CounterTerm;\Lambda^2}\Big|_{SymmetryBreaking}$ in Eq. (2A.1), an exhaustive search is done for $\langle H_{Bare} \rangle$-dependent terms in Feynman graphs arising in the **broken** theory from $L_{L\Sigma M}^{Bare;\Lambda^2}$ (which has $\langle H \rangle \neq 0$) in Eq. (2A.1): these are written explicitly as Taylor series in $\langle H_{Bare} \rangle$;

- The terms in the **broken** theory's Taylor series are demonstrated to be "…what one gets if one evaluates the amplitude for emission of *s* pairs of *H*'s from a pion (*H*) line which carries the initial momentum *k*, and then let all the momenta of the emitted *H*'s go to zero", Lee in Ref. [11]. A diagrammatic proof and interpretation of this fact, including proof that all associated (e.g. external leg, time-ordered product) symmetry factors are correct, is constructed;

- Remembering all-orders renormalization of the **symmetric** theory, Lee writes: "… if we renormalize the $H_{Bare}$ field and its vacuum expectation value $\langle H_{Bare} \rangle$ as well as the $\vec{\pi}$ field of the (**broken** $O(4)L\Sigma M$) … according to

$$(H, \vec{\pi})^{Bare} = Z_\Phi^{1/2} (H, \vec{\pi})^{\text{Re}normalized};$$
$$\lambda_{\text{Re}normalized}^2 = Z_\Phi^{1/2} Z_{\lambda^2} \lambda_{Bare}^2; \quad (A4.2)$$

and (a crucial observation for this paper) B.W.Lee's Eq. (4b.2) in Ref. [11]

$$\langle H \rangle^{Bare} = Z_\Phi^{1/2} \langle H \rangle^{\text{Re}normalized}; \quad (A4.3)$$

where $Z_\Phi$ is the wave function renormalization constant of the **symmetric** theory, then the renormalized expression for (the before-mentioned Taylor series in $\langle H_{Bare} \rangle$ plus included counter-terms in the **broken** theory) is also finite in terms of $\lambda^2_{Renormalized}$ and $\langle H_{Renormalized} \rangle$" [11];

- Lee proves that all S-Matrix UV-QD in the **broken** theory are absorbed into $\mu^2_{Bare}$: i.e. proving, for this paper, that they can be absorbed in $\delta\mu^2$ in Eq. (2A.1);

- Lee then re-includes $L^{CounterTerm;\Lambda^2}_{L\Sigma M}\big|_{SymmetryBreaking}$ in the **broken** theory, first proving the Ward identity

$$Z^{1/2}_\Phi \varepsilon_{L\Sigma M} = \langle H \rangle^{Renormalized} m^2_\pi; \tag{A4.4}$$

where $m^2_\pi$ is the physical renormalized pseudo-scalar pion (pole) mass-squared. Using Eq. (A4.2, A4.3, A4.4), he proves the finiteness of the eigenvalue equation for $\langle H_{Bare} \rangle$

$$\varepsilon_{L\Sigma M} - \langle H_{Bare} \rangle \left( \mu^2_{Bare} + \lambda^2_{Bare} \langle H_{Bare} \rangle^2 \right) - S\left(\langle H_{Bare} \rangle\right) = 0; \tag{A4.5}$$

where $-S(\langle H \rangle_{Bare})$ are the higher-order contributions to the Higgs VEV, i.e. tadpole diagrams. This eigenvalue equation is simpler than it looks: e.g. at 1-loop it can be re-written (including all UV-QD) $\varepsilon_{L\Sigma M} - \langle H \rangle m^2_\pi = 0$, and is interpreted as the Ward-Takahashi identity Eq. (A4.4), and the tautology Eq. (2B.6).

**Appendix 5: Proof that UV-QD do not arise in the 1-loop renormalization of mass-less 4-fermion high precision electroweak (HPEW) processes**

We begin by explicit calculation (agreed long ago [2,3,4,5,6,7,8,9]) of 1-loop $(Z_\mu, W^\pm_\mu)$ 2-point self-energies in the SM, i.e. the so-called "oblique" [18] corrections. Notation here [5]: $B_0(q^2, M^2_W, M^2_Z) = B_0(W,Z)$ and $(Q_f, I_{3;f})$ are the electric charge and 3rd component of isospin for a virtual fermion. For 3 generations

$$W = M^2_W; \quad Z = M^2_Z; \quad H = m^2_h; \quad 0 = m^2_{Photon};$$

$$f = m^2_{fermion}; \quad u = m^2_{u_{Quark}}; \quad d = m^2_{d_{Quark}}; \quad v = m^2_{NeutralLepton}; \quad e = m^2_{ChargedLepton}; \tag{A5.1}$$

$$u_{Quark} = u,c,t; \quad d_{Quark} = d,s,b; \quad v = v_e, v_\mu, v_\tau; \quad e = e, \mu, \tau;$$

For simplicity and pedagogical clarity, we address only **"oblique loop"** [5,18] effects, excluding (finite after renormalization) gauge invariant vertex and box contributions, SM baroque-ness (e.g. GIM) and certain imaginary parts: all these are written in Ref. [5]. The oblique contributions from virtual SM bosons to $(Z_\mu, W^\pm_\mu)$ 2-point self-energies are displayed in $R_\xi : \xi = 1$ gauge [5]:

$$16\pi^2\Pi_{33}(q^2) = q^2\left[-9B_3(W,W) + \frac{7}{4}B_0(W,W) + \frac{2}{3}\right] - 2M_W^2 B_0(W,W)$$

$$+ q^2\left[-B_3(Z,H) - \frac{1}{4}B_0(Z,H)\right] + \frac{1}{4}\left(M_Z^2 - M_H^2\right)\left[2B_1(Z,H) + B_0(Z,H)\right] \quad \text{(A5.2)}$$

$$+ M_Z^2 B_0(Z,H);$$

$$16\pi^2\Pi_{3Q}^T(q^2) = q^2\left[-10B_3(W,W) + \frac{3}{2}B_0(W,W) + \frac{2}{3}\right];$$

$$16\pi^2\Pi_{3Q}^L(q^2) = -2M_W^2 B_0(W,W);$$

$$16\pi^2\Pi_{QQ}(q^2) = q^2\left[-12B_3(W,W) + B_0(W,W) + \frac{2}{3}\right];$$

$$16\pi^2\Pi_{+-}(q^2) = \frac{2}{3}q^2 + s_\theta^2\left(q^2\left[-8B_3(W,0) + 2B_0(W,0)\right] + 2M_W^2\left[2B_1(W,0) + B_0(W,0)\right]\right)$$

$$+ c_\theta^2\left(q^2\left[-8B_3(W,Z) + 2B_0(W,Z)\right] + 2\left(M_W^2 - M_Z^2\right)\left[2B_1(W,Z) + B_0(W,Z)\right]\right)$$

$$+ q^2\left[-B_3(W,Z) - \frac{1}{4}B_0(W,Z)\right] + \frac{1}{4}\left(M_W^2 - M_Z^2\right)\left[2B_1(W,Z) + B_0(W,Z)\right]$$

$$+ \left(M_Z^2 - 3M_W^2\right)B_0(W,Z)$$

$$+ q^2\left[-B_3(W,H) - \frac{1}{4}B_0(W,H)\right] + \frac{1}{4}\left(M_W^2 - M_H^2\right)\left[2B_1(W,H) + B_0(W,H)\right]$$

$$+ M_W^2 B_0(W,H);$$

The oblique contributions from virtual SM fermions to $(Z_\mu, W_\mu^\pm)$ 2-point self-energies are [5]:

$$16\pi^2\Pi_{33}(q^2) = \sum_{Quarks} N_{Colors} I_{3;f}^2\left[4q^2 B_3(f,f) - 2m_f^2 B_0(f,f)\right]$$

$$+ \sum_{Leptons} I_{3;f}^2\left[4q^2 B_3(f,f) - 2m_f^2 B_0(f,f)\right];$$

$$16\pi^2\Pi_{3Q}^T(q^2) = q^2\sum_{Quarks} 4N_{Colors} Q_f I_{3;f} B_3(f,f) + q^2\sum_{Leptons} 4Q_f I_{3;f} B_3(f,f);$$

$$16\pi^2\Pi_{3Q}^L(q^2) = 0; \quad \text{(A5.3)}$$

$$16\pi^2\Pi_{QQ}(q^2) = q^2\sum_{Quarks} 8N_{Colors} Q_f^2 B_3(f,f) + q^2\sum_{Leptons} 8Q_f^2 B_3(f,f);$$

$$16\pi^2\Pi_{+-}(q^2) = \sum_{QuarkDoublets} N_{Colors}\left\{2q^2 B_3(u,d) + m_u^2 B_1(d,u) + m_d^2 B_1(u,d)\right\}$$

$$+ \sum_{LeptonDoublets}\left\{2q^2 B_3(\nu,e) + m_\nu^2 B_1(e,\nu) + m_e^2 B_1(\nu,e)\right\};$$

The crucial observation is that **1-loop UV-QD do not arise in** $\Pi_{33}, \Pi_{3Q}^T, \Pi_{3Q}^L, \Pi_{QQ}, \Pi_{+-}$ **and oblique radiative corrections.** That is, $B_0, B_1, B_{21}, B_3 = B_{21} + B_1$ in Eqs. (A1.3, A1.1, A5.3, A5.4) are only, at worst, logarithmically divergent. Any UV-QD appearing in intermediate steps of the calculations have cancelled exactly using self-consistency relations (e.g. change of integration variables) such as in Eq. (A1.4, A1.5): a remnant of that exact self-consistency cancellation is the appearance of $B_{21}$, but not $B_{22}$ or $A$, in Eq. (A5.2, A5.3). **Eqs. (A5.2) and**

**(A5.3) prove that UV-QD do not arise in 1-loop SM $(Z_\mu, W_\mu^\pm)$ 2-point functions or high precision electroweak (HPEW) physics!**

We turn to renormalization of the SM HPEW processes. The Higgs VSC, Eq. (3B.8), first eliminates all tadpole contributions to fermion and boson lines. The bare $SU(2)$ and $U(1)$ gauge couplings, Higgs VEV and $\rho_{EJV}$, ($g_{Bare}, g'_{Bare}, \langle H_{Bare} \rangle, \rho_{Bare}$), are then related to their $(q^2)$-dependent renormalized values (including threshold effects), the UV finite so-called "Star" renormalization scheme" [5,1], by

$$\frac{1}{g_*^2(q^2)} = \frac{1}{g_{Bare}^2} - \text{Re}\left[\frac{\Pi_{3Q}^T(q^2)}{q^2} - 2\frac{\Pi_{3Q}^L(q^2)}{M_W^2}\right];$$

$$\frac{1}{\left(g'_*(q^2)\right)^2} = \frac{1}{\left(g'_{Bare}\right)^2} - \text{Re}\left[\frac{\Pi_{QQ}(q^2)}{q^2} - \frac{\Pi_{3Q}^T(q^2)}{q^2}\right]; \quad (A5.4)$$

$$\frac{1}{4\sqrt{2}G_\mu^*(q^2)} = \frac{1}{2}\langle H_{Bare}\rangle^2 - \text{Re}\left[\Pi_{+-}(q^2) - \Pi_{3Q}^T(q^2) - 2\Pi_{3Q}^L(q^2)\right];$$

$$\frac{1}{4\sqrt{2}G_\mu^*(q^2)\rho_*(q^2)} = \frac{1}{2}\langle H_{Bare}\rangle^2 - \text{Re}\left[\Pi_{33}(q^2) - \Pi_{3Q}^T(q^2) - 2\Pi_{3Q}^L(q^2)\right]; \quad (A5.5)$$

where the longitudinal $\Pi_{3Q}^L(q^2)$ terms ensure gauge invariance [5,16,1]. **Since $\rho_*(q^2)$ is finite, SM <u>electro-weak gauge sector</u> bare or renormalized parameters have, at 1-loop, to absorb at worst logarithmic divergences: there are no UV-QD in Eqs. (A5.4,A5.5).**

The Star Scheme proved useful in practice for comparing all different HPEW experiments with each other [5]. A single neutral current, and a single charged current, effective matrix element for all mass-less 4-fermion HPEW processes (and Dyson and renormalization group re-summation of all 1-loop oblique loops) form a running and threshold improved Born approximation [5], ideal for HPEW experimental data fitting [9, D.Levinthal *et. al* in Ref. 8].

$$M_{NeutralCurrent}^{ObliqueHPEW}(q^2) = e_*^2 \frac{QQ'}{q^2} + \frac{e_*^2}{s_*^2 c_*^2} \frac{(I_3 - Qs_*^2)(I_3' - Q's_*^2)}{q^2 + \frac{e_*^2}{s_*^2 c_*^2}\frac{1}{4\sqrt{2}G_\mu^* \rho_*} - i\sqrt{s}\Gamma_Z^*};$$

$$M_{Ch\arg edCurrent}^{ObliqueHPEW}(q^2) = \frac{e_*^2}{s_*^2}\frac{I_+ I_-' + I_- I_+'}{q^2 + \frac{e_*^2}{s_*^2}\frac{1}{4\sqrt{2}G_\mu^*} - i\sqrt{s}\Gamma_W^*}; \quad (A5.6)$$

$$\sqrt{s}\Gamma_Z^* = \frac{e_*^2}{s_*^2 c_*^2}\text{Im}\left[\Pi_{33} + 2s_*^2 \Pi_{3Q} + s_*^4 \Pi_{QQ}\right]; \qquad \sqrt{s}\Gamma_W^* = \frac{e_*^2}{s_*^2}\text{Im}\left[\Pi_{+-}\right];$$

For pedagogical clarity, only the $(Z_\mu, W_\mu^\pm)$ widths are shown in Eq. (A5.6): all the other imaginary parts, necessary for full matrix element unitarity, are found in Ref. [5]. Because all oblique corrections to HPEW processes had been so classified, all non-decoupling effects of virtual heavy particles in oblique loops (i.e. SM particles $m_{TopQuark}^2, m_{Higgs}^2$ and Beyond the Standard Model (BSM) particles) could also be classified by experimental sensitivity [5] in Eq. (A5.6). Ref. [5] first showed that all non-decoupling effects (i.e. of very heavy virtual SM and BSM particles) can affect HPEW in only three combinations [5]: **dimension-2** $\Delta_\rho(0)$ and **dimension-zero** $\Delta_3/q^2, \Delta_+/q^2$

$$\Delta_+(q^2) = \frac{1}{4\sqrt{2}G_\mu^*(0)} - \frac{1}{4\sqrt{2}G_\mu^*(q^2)}$$
$$= \text{Re}\left[\Pi_{+-}(q^2) - \Pi_{3Q}^T(q^2) - 2\Pi_{3Q}^L(q^2)\right] - \text{Re}\left[\Pi_{+-}(0) - 2\Pi_{3Q}^L(0)\right];$$

$$\Delta_3(q^2) = \frac{1}{4\sqrt{2}G_\mu^*(0)\rho_*(0)} - \frac{1}{4\sqrt{2}G_\mu^*(q^2)\rho_*(q^2)}$$
$$= \text{Re}\left[\Pi_{33}(q^2) - \Pi_{3Q}^T(q^2) - 2\Pi_{3Q}^L(q^2)\right] - \text{Re}\left[\Pi_{33}(0) - 2\Pi_{3Q}^L(0)\right];$$

$$\Delta_\rho(0) = \text{Re}\left[\Pi_{33}(0) - \Pi_{+-}(0)\right] = \frac{1}{4\sqrt{2}G_\mu^*(0)}\left(1 - \frac{1}{\rho_{EJV}}\right) = \frac{\alpha_{QED}}{4\sqrt{2}G_\mu^*(0)}T;$$

(A5.7)

$$\frac{\Delta_3(q^2)}{q^2} \xrightarrow{q^2 \to 0} -\frac{1}{16\pi}S;$$

$$\frac{\Delta_+(q^2)}{q^2} \xrightarrow{q^2 \to 0} \frac{1}{16\pi}(-S+U);$$

(A5.8)

Eq. (A5.8) gives the 1-to-1 relationship with modern *T, S* and U [17] and ancient $\rho_{EJV}$.

The $(\alpha, G_\mu^{Experimental}, M_Z^2)$ renormalization scheme, universally established and used throughout HPEW LEP1/SLC physics, and first proposed in Ref. [18], fixes the physical values of Thomson scattering, the muon decay lifetime and the $Z^0$ mass. Defining the renormalized running electric charge and weak mixing angle

$$\frac{1}{e_*^2(q^2)} = \frac{1}{g_*^2(q^2)} + \frac{1}{\left(g_*^\prime(q^2)\right)^2}; \quad s_*^2(q^2) = \frac{e_*^2(q^2)}{g_*^2(q^2)}; \quad s_*^2(q^2) + c_*^2(q^2) = 1;$$ (A5.9)

the physical scales, for four-massless fermion HPEWP processes mediated by the gauge sector SM, are set using the experimental input data

$$\frac{e_*^2(0)}{4\pi} = \alpha_{QED} = \frac{1}{137.03602}; \qquad G_\mu^*(0) = G_\mu^{Experimental} = 1.16637x10^{-5} GeV^{-2};$$

$$\left[\frac{e_*^2}{s_*^2 c_*^2}\frac{1}{4\sqrt{2}G_\mu^*\rho_*}\right]_{q^2=-M_Z^2} = M_Z^2 = (91.1876\pm0.0021 \text{ GeV})^2;$$

(A5.10)

together with calculation of $\rho_*(q^2 = -M_Z^2)$. The resulting HPEW predictions [2,3,4,5,6,7,8,9] of the full SM, facilitated by the <u>running</u> improved Born approximation [5], have been tested and vindicated to better than 1% accuracy by LEP1/SLC [9] and other HPEW experiments.

**<u>Appendix 6</u>: The SM (almost certainly) has exactly zero remnant of 1PI multi-loop UV-QD and does not suffer a HFTP at <u>any</u> loop-order of electro-weak and QCD perturbation theory**

The SM was renormalized to all perturbative loop-orders long ago: we remind the reader of results relevant to this paper [19,20,21,11,12] here. We follow closely the reasoning in Sections 2D, 3C and Ref. [10].

It is convenient to work in "Goldstone gauge", by adding to the gauge-invariant SM bare Lagrangian (which includes fermions, gauge bosons, the linear representation of the $\Phi$ doublet and the Higgs mechanism), gauge-fixing and ghost terms more usually associated with pure gauge theories (which would only include gauge bosons);

$$L_{SM;GaugeFixing+Ghost}^{Bare;GoldstoneGauge} = L_{SU3Color;GaugeFixing+Ghost}^{Bare}$$
$$+ L_{SU2Isospin;GaugeFixing+Ghost}^{Bare} + L_{U1Hypercharge;GaugeFixing+Ghost}^{Bare}$$

$$L_{SU3Color;GaugeFixing+Ghost}^{Bare} = -\frac{1}{2\xi_{SU3}}\left(\partial_\mu G_\mu^A\right)^2 + \bar{\xi}^A \partial_\mu D_\mu^{AB} \eta^B;$$

$$L_{SU2Isospin;GaugeFixing+Ghost}^{Bare} = -\frac{1}{2\xi_{SU2}}\left(\partial_\mu W_\mu^a\right)^2 + \bar{\xi}^a \partial_\mu D_\mu^{ab} \eta^b;$$

(A6.1)

$$L_{U1Hypercharge;GaugeFixing+Ghost}^{Bare} = -\frac{1}{2\xi_{U1}}\left(\partial_\mu B_\mu\right)^2 + \bar{\xi}\partial_\mu D_\mu \eta;$$

This reduces the local gauge symmetries to global BRST symmetries [21]. The purpose of Goldstone gauge is to decouple the ghosts from the scalars and make manifest that there are no UV-QD in the gauge-ghost-fermion sector of the all-loop UV-QD SM Lagrangian $L_{SM}^{All-Loop;\Lambda^2}$. We group all of the bosons together

$$\Pi = G_\mu^A, W_\mu^a, B_\mu; \bar{\xi}^A, \eta^A, \bar{\xi}^a, \eta^a, \bar{\xi}, \eta; H, \pi^a;$$ (A6.2)

and construct the UV-QD part $L_{SM;\Phi-Sector}^{All-Loop;\Lambda^2}$ of the all-loop-order SM effective Lagrangian $L_{SM;\Phi-Sector}^{Effective;All-Loop;\Lambda^2}$ from global BRST-invariant operators and gauge-dependent tadpoles. The coefficients of the relevant operators arising from 1PI multi-loop calculations have superficial degree of divergence:

$$D_{SM} = 4 - N_\Pi - \frac{3}{2}N_\Psi - N_\partial - V_{\Pi\Pi\Pi};$$ (A6.3)

- $N_\Pi$ is the number of external $\Pi$ boson legs Eq. (A6.2);
- $N_\Psi$ is the number of external $\Psi$ fermion legs;
- $N_\partial$ is the number of derivatives acting on external legs;
- $V_{\bar{\Psi}\Psi\Pi}$, the number of internal fermion-boson vertices, does not appear in $D_{SM}$;
- $V_{\Pi\Pi\Pi}$ is the number of internal 3-$\Pi$ boson vertices;
- $V_{\Pi\Pi\Pi\Pi}$, the number of internal 4-$\Pi$ boson vertices, does not appear in $D_{SM}$;
- $V_{\partial\Pi\Pi\Pi}$, the number of internal once-differentiated 3-$\Pi$ boson vertices, does not appear in $D_{SM}$;
- $V^{CounterTerm;\Lambda^2}$, the number of internal UV-QD counter-term insertions, does not appear in $D_{SM}$ when it is assumed that $\delta\mu^2 \sim \Lambda^2, \varepsilon_{SM}^\xi \sim \Lambda^2 \langle H \rangle$ for the case of UV-QD. Such insertions appear, e.g. inside 1PI 1-loop diagrams in order to cancel UV-QD nested within 1PI 2-loop diagrams;

We are only interested in 1PI multi-loop UV-QD in the coefficients of relevant dimension-2 operators. $N_\Pi = 0$ vacuum energy/bubbles are ignored in this paper. In the zero-external-ghost unbroken $N_\Pi = 2$ sector, some 2-point functions are embedded in BRST invariant $D_{SM} = 0$ operators at worst logarithmically divergent:

$$\left(G_{\mu\nu}^{A}\right)^{2},\left(W_{\mu\nu}^{a}\right)^{2},\left(B_{\mu\nu}\right)^{2},\left|D_{\mu}\Phi\right|^{2} \qquad (A6.4)$$

In Goldstone gauge, ghosts can only appear married to gauge fields in BRST-invariant combinations, which are also at worst logarithmically divergent and proportional to one of

$$L_{SU3Color;GaugeFixing+Ghost}^{Bare}, L_{SU2Isospin;GaugeFixing+Ghost}^{Bare}, L_{U1Hypercharge;GaugeFixing+Ghost}^{Bare}; \qquad (A6.5)$$

in Eq. (A6.1). The SM therefore has no UV-QD outside the $\Phi$ sector. This implies that UV-QD do not arise in any loop order in 4-fermion high precision electroweak (HPEW) processes, where the Yukawa couplings of the four external fermions are zero. Since $G_{\mu}^{Experimental}\langle H_{Bare}\rangle^{2}$ is renormalized by such a HPEW process (i.e. muon lifetime), we conclude that $\langle H \rangle$ absorbs no UV-QD, only at worst logarithmic divergences, and is not fine-tuned.

In the $\Phi$ sector, UV-QD can arise from
- $N_{\Pi}=1$: gauge-dependent 1PI loop tadpoles with at least one 3-boson vertex $V_{\Pi\Pi\Pi}$;
- $N_{\Pi}=2$: BRST invariant 2-point scalar-sector functions proportional to $\Phi^{\dagger}\Phi$;

But the SM bare Lagrangian counterterm Eq. (3B.9)

$$L_{SM;\Phi-Sector}^{CounterTerm;\Lambda^{2}} = -\delta\mu^{2}\left[\frac{1}{2}h^{2}+\frac{1}{2}\pi_{3}^{2}+\pi_{+}\pi_{-}+\langle H\rangle h\right]+\varepsilon_{SM}^{\xi}H; \qquad (A6.6)$$

is guaranteed to remove all 1PI UV-QD in the SM S-Matrix: i.e. the $D_{SM}=2$ contributions with $N_{\Pi}=2$ and $N_{\Pi}=1$ [19,20,21,11,12]. Excluding logarithmic divergences and finite parts, the 1PI multi-loop UV-QD contribution can therefore be written:

$$L_{SM}^{All-Loop;\Lambda^{2}} = C_{SM}^{All-loop;\Lambda^{2}}\Lambda^{2}\left(\frac{1}{2}h^{2}+\frac{1}{2}\pi_{3}^{2}+\pi_{+}\pi_{-}+C_{SM;\xi}^{All-loop;\Lambda^{2}}\langle H\rangle h\right); \qquad (A6.7)$$

with $C_{SM}^{All-loop;\Lambda^{2}}$ and $C_{SM;\xi}^{All-loop;\Lambda^{2}}$ finite constants dependent (i.e. because of nested divergences within multi-loop 1PI graphs) on the finite physical input constant parameters of the theory.

We form the all-loop-UV-QD-improved effective Lagrangian for the SM scalar sector, including all-orders scalar 2-point self-energy and 1-point tadpole UV-QD (but ignoring logarithmically divergent and finite contributions and vacuum energy/bubbles)

$$L_{SM;\Phi-Sector}^{Effective;;All-Loop;\Lambda^{2}} = L_{SM;\Phi-Sector}^{Bare;\Lambda^{2}} + L_{SM}^{All-Loop;\Lambda^{2}}$$
$$= -\left|D_{\mu}\Phi\right|^{2} - V_{SM}^{\mathrm{Re}normalized;All-Loop;\Lambda^{2}}; \qquad (A6.8)$$

$$V_{SM}^{\mathrm{Re}normalized;All-Loop;\Lambda^{2}} = \lambda^{2}\left[\frac{1}{2}h^{2}+\frac{1}{2}\pi_{3}^{2}+\pi_{+}\pi_{-}+\langle H\rangle h\right]^{2}$$
$$+\left(\delta\mu^{2}-C_{SM}^{All-Loop;\Lambda^{2}}\Lambda^{2}\right)\left[\frac{1}{2}h^{2}+\frac{1}{2}\pi_{3}^{2}+\pi_{+}\pi_{-}\right] \qquad (A6.9)$$
$$+\left[\left(\delta\mu^{2}-C_{SM;\xi}^{All-Loop;\Lambda^{2}}C_{SM}^{All-Loop;\Lambda^{2}}\Lambda^{2}\right)\langle H\rangle-\varepsilon_{SM}^{\xi}\right]h;$$

but it is not gauge invariant! Now impose the two SM GMRP conditions:
- <u>Higgs VSC</u>: the Higgs must not simply disappear into the vacuum;
$$\varepsilon_{SM}^{\xi}-\left(\delta\mu^{2}-C_{SM;\xi}^{All-Loop;\Lambda^{2}}C_{SM}^{All-Loop;\Lambda^{2}}\Lambda^{2}\right)\langle H\rangle=0; \qquad (A6.10)$$
- <u>Goldstone SRC</u>: $m_{\pi;NGB;SM}^{2}$ is a 1PI multi-loop-induced artefact, the solution to a highly non-linear equation which, after tadpole renormalization, has absorbed all remaining SM UV-QD to all perturbative loop-orders. **The Goldstone SRC insist that the SM NGB mass-squared $m_{\pi;NGB;SM}^2$ must be exactly zero**:

$$m_{\pi;NGB;SM}^2 = \left(\delta\mu^2 - C_{SM}^{All-Loop;\Lambda^2}\Lambda^2\right) = 0; \tag{A6.11}$$

**Gauge invariance is then restored to the scalar sector SM effective Lagrangian**

$$L_{SM;\Phi-Sector}^{Effective;All-Loop;\Lambda^2} = -\left|D_\mu\Phi\right|^2 - V_{SM}^{Renormalized;All-Loop;\Lambda^2};$$

$$V_{SM}^{Renormalized;All-Loop;\Lambda^2} = \lambda^2\left[\frac{1}{2}h^2 + \frac{1}{2}\pi_3^2 + \pi_+\pi_- + \langle H\rangle h\right]^2 = \lambda^2\left(\Phi^\dagger\Phi - \frac{1}{2}\langle H\rangle^2\right)^2; \tag{A6.12}$$

A crucial result is that all-loop-orders SM UV-QD contributions to the Higgs' mass-squared in Eq. (A6.9 and A6.11) also vanish identically:

$$\left(\delta\mu^2 - C_{SM}^{All-Loop;\Lambda^2}\Lambda^2\right)\left(\frac{1}{2}h^2\right) = m_{\pi;NGB;SM}^2\left(\frac{1}{2}h^2\right) = 0; \tag{A6.13}$$

The UV-corrected effective SM scalar sector Lagrangian Eq. (A6.12) (which includes all perturbative electro-weak and QCD 1PI UV-QD multi-loop contributions, but ignores logarithmic divergences, finite contributions and vacuum energy/bubbles), therefore gives the sensible, at worst logarithmically divergent and not-fine-tuned Higgs mass-squared:

$$L_{SM;\Phi-Sector}^{Effective;All-Loop;\Lambda^2} = -\frac{1}{2}m_h^2 h^2 + ++; \quad m_h^2 = 2\lambda^2\langle H\rangle^2; \tag{A6.14}$$

The reader is warned that **Appendix 6 does not constitute a rigorous quantum field theoretic proof** that all SM S-Matrix UV-QD vanish to all perturbative QCD and electroweak loop orders. Rigorous mathematical proof at the level of all-loop-orders nested divergences is required. Among the issues:

- $L_{SM;\Phi-Sector}^{Effective;1-Loop;\Lambda^2} \xrightarrow[g_1,g_2,g_3 \to 0]{} L_{GoldstoneMode;L\Sigma M+SMQuarks\&Leptons;\Phi-Sector}^{Effective;1-Loop;\Lambda^2}$
- $\langle H_{Bare}\rangle$ doesn't absorb UV-QD during HPEW renormalization to $G_\mu^{Experimental}$;
- Full nested divergence disentanglement of gauge-dependent UV-QD tadpole renormalization Eq. (A6.10). Determination of whether it leaves un-physical 1PI loop-induced artefact parameters, e.g. $m_{\pi;NGB;SM}^2$, gauge invariant;
- Proof that all physical input parameters (e.g. $m_h^2, \lambda^2, \langle H\rangle^2$) are gauge invariant after unphysical NGB masses $m_{\pi;NGB;SM}^2 \to 0$;
- Proof that all 1PI loop-induced-artefact unphysical operators (e.g. $\langle H\rangle m_{\pi;NGB;SM}^2 H$ in Eq. (3B.12)) vanish when $m_{\pi;NGB;SM}^2 \to 0$;
- Treatment of infra-red effects;